\documentclass[12pt,a4paper]{article} 
\usepackage{graphicx} 
\usepackage{amsmath} 
\usepackage{amssymb} 
\usepackage{slashed} 
\usepackage{bigstrut} 
\usepackage{stmaryrd} 
\usepackage{mathtools} 
\usepackage{multirow} 

\newcommand{\beq}{\begin{equation}} 
\newcommand{\eeq}{\end{equation}} 
\def\bsp#1\esp{\begin{split}#1\end{split}} 
\def\bal#1\eal{\begin{align}#1\end{align}} 
\newcommand{\beeq}{\begin{eqnarray}} 
\newcommand{\eeeq}{\end{eqnarray}}

\newcommand{\pythia}   {\texttt{PYTHIA}}

\newcommand{\powhel}   {\texttt{PowHel}} 
 
\newcommand{\powhegbox}{\texttt{POWHEG-BOX}} 
 
\newcommand{\helacnlo} {\texttt{HELAC-NLO}}

\newcommand{\helaconeloop} {\texttt{HELAC-1LOOP}} 
 
\newcommand{\helacdipole} {\texttt{HELAC-Dipole}}

\newcommand{\pt}       {\ensuremath{p_\bot}} 
 
\newcommand{\kt}       {\ensuremath{k_{\bot}}} 
\newcommand{\pTmiss}{\ensuremath{\slash\hspace*{-5pt}{p}_{\perp}}}

\newcommand{\bq}       {\ensuremath{\mathrm{b}}}

\newcommand{\bB}       {\ensuremath{\mathrm{b\,\bar{b}}}} 
\newcommand{\tT}       {\ensuremath{\mathrm{t\,\bar{t}}}}

\newcommand{\res}[3]{\ensuremath{{#1}^{#2}_{#3}}} 
\newcommand\fig[1]     {Fig.\,{\ref{#1}}} 
\newcommand\figs[2]    {Figs.\,{\ref{#1}} and ~\ref{#2}} 
\newcommand\figss[2]   {Figs.\,{\ref{#1}}--\ref{#2}} 
\newcommand\sect[1]    {Sect.\,{\ref{#1}}}

\newcommand{\ttbb}     {\ensuremath{\mathrm{t \bar{t} b \bar{b}}}}

\newcommand{\ttjj}     {\ensuremath{\mathrm{t \bar{t} j j}}} 
\newcommand{\ttH}      {\ensuremath{\mathrm{t \bar{t}} H}}

\begin{document} 
%\maketitle 
%\flushbottom 
 
\begin{titlepage} 
\vspace*{-2cm} 
\vskip .5in 
\begin{center} 
{\large\bf 
Hadroproduction of \ttbb\ final states at LHC:\\ 
predictions at NLO accuracy matched with Parton Shower 
}\\ 
\vspace*{1.5cm} 
{\large M.~V.~Garzelli, A.~Kardos and Z.~Tr\'ocs\'anyi} \\ 
\vskip 0.2cm 
MTA-DE Particle Physics Research Group and Institute of Physics,\\ 
  University of Debrecen, H-4010 Debrecen P.O.Box 105, Hungary 
\vskip 1cm 
\end{center} 
 
\par \vspace{2mm} 
\begin{center} {\large \bf Abstract} \end{center} 
\begin{quote} 
We present predictions for the hadroproduction of \ttbb\ final states at 
the LHC with collision energies $\sqrt{s} = 8$\,TeV and 14\,TeV 
at NLO accuracy matched with parton shower, as obtained with \powhel + 
\pythia.  We quantify the effects of parton shower and hadronization.  
We find these are in general moderate except the effect of the decay of 
heavy particles, which can modify significantly some distributions, 
like that of the invariant mass of the two leading b-jets. 
We also show kinematic distributions obtained with cuts inspired to 
those recently employed by the CMS collaboration. For these predictions,  
we present the theoretical uncertainty bands, related to both scale and 
PDF variations. We find that these uncertainties are only moderately 
affected by the change of the collision energy. 
\end{quote}  
 
\vspace*{\fill} 
\begin{flushleft} 
July 2014 
\end{flushleft} 
\end{titlepage} 
 
\section{Introduction} 
\label{intro} 
 
The \ttbb\ hadroproduction process was one of the first four-parton 
hadroproduction processes studied a few years ago in order to test 
and show the capabilities of new unitarity-inspired methods for 
evaluating one-loop amplitudes, as well as the potential of new 
developments in the more traditional tensor-reduction approach. 
Pioneering works in this direction were performed  by Bredenstein 
{\it et al.}~\cite{Bredenstein:2009aj,Bredenstein:2010rs}, as 
well as by the \helacnlo\ collaboration~\cite{Bevilacqua:2009zn}.  
In both cases the role of next-to-leading-order (NLO) QCD corrections was 
shown, by considering rather loose systems of cuts. The agreement 
between predictions obtained by different methods was taken as a prove 
of the correctness of the new conceptual developments, that were 
subsequently applied successfully to other multiparticle hadroproduction 
processes, reaching a level of unforeseen complexity \cite{Bevilacqua:2010ve,Cullen:2013saa,Alwall:2014hca,Bern:2014voa}. 
 
A further application was represented by the study of \ttbb\ production
at the TeVatron~~\cite{Worek:2011rd}. 
Nowadays, a renewed interest for this process has arisen, driven not only 
by further theoretical developments, including the possibility of 
matching complex NLO QCD calculations to parton shower (PS) approaches, 
but also by the requests of the experimental collaborations working at 
LHC, that need input to reduce the uncertainties on the estimate of the 
\ttbb\ contribution to the total background in their searches for the 
Higgs boson in the \ttH\ channel, with the H boson decaying in a \bB\ 
pair~\cite{ATLAS-CONF-2012-135,Chatrchyan:2013yea,CMS-PAS-HIG-13-019}. 
The theoretical uncertainty quoted in recent experimental analyses on the 
contribution of this irreducible non-resonant background component amounts to a 
few tens percent ($\sim$ 50 \% in Ref. ~\cite{CMS-PAS-HIG-13-019}). We expect 
that theoretical estimates with the best possible theoretical accuracy 
within present reach of the most advanced event generators can help 
reduce this uncertainty. Thus, one of the aims of this paper is to present predictions for differential cross sections at NLO QCD accuracy matched to PS and hadronization, resulting in predictions at the hadron level, using cuts similar to 
those employed by the experimental collaborations.   
 
First predictions concerning  \ttbb\ computations with NLO QCD accuracy 
matched with PS were presented by our group in 
Ref.~\cite{Kardos:2013vxa,Heinemeyer:2013tqa}, making use of \powhel. 
More recently, \ttbb\ predictions obtained by using the MC@NLO matching 
algorithm~\cite{Frixione:2002ik}, as encoded in {\texttt{SHERPA}} 
\cite{Gleisberg:2008ta}, and massive b-quarks, with a fixed pole mass 
$m_\bq = 4.75$\,GeV, have been reported in Ref.~\cite{Cascioli:2013era}, 
using 1-loop amplitudes computed by {\texttt{OpenLoops}}~\cite{Cascioli:2011va}. 
Our computation differs from that presented in Ref.~\cite{Cascioli:2013era} in the following aspects: we use (i) massless b-quarks in the generation of NLO matrix elements, (ii) a different matching algorithm (POWHEG~\cite{Nason:2004rx,Frixione:2007vw}), and (iii) a different Shower Monte Carlo program (\pythia).  
 
As pointed out in Ref.~\cite{Cascioli:2013era}, the finite mass of the 
b-quarks allows for computing the contribution of \ttjj\ events with 
the jets originating from two gluons that both separately split into a 
collinear \bB\ pair.  Each pair itself becomes part of one $b$-jet (as 
the b-quarks are sufficiently collinear to be unresolved by the jet 
algorithm), so the final state contains a \tT-quark pair in association 
with two $b$-jets. Strictly speaking, such contributions are higher 
order in perturbation theory
and are not included in our computation with NLO+PS accuracy, 
where collinear $b\bar{b}$ pairs can only be produced by gluon splitting 
in the PS. On the other hand, in Ref.~\cite{Cascioli:2013era}, one of those 
$g\to \bB$ splittings is included in the real radiation matrix elements 
(at LO accuracy for these contributions), while the other is generated 
by the parton shower, and such double $g\to \bB$ splittings were found 
to give a significant contribution to the \tT+2 $b$-jet final states when 
the t-quarks are kept stable.  
  
This paper represents an extension of our previous studies, relying on 
a more advanced \powhel\ implementation, on the basis of recent 
\powhegbox\ developments, and includes a detailed study of the 
theoretical uncertainties affecting our results in the most general 
case, including top decay and the full SMC chain up to hadronization  
and hadron decay.   
 
\iffalse 
The paper is organized as follows: the methods applied in the theoretical 
calculation are summarized in \sect{sec:method}, followed by 
predictions under different sets of cuts in \sect{sec:phenomenology}. 
A discussion on the uncertainties on these predictions, due to different 
sources, is also presented in \sect{sec:phenomenology}, and finally 
our conclusions on the present status of the calculation and on possible 
further developments, especially as for the uncertainty estimates, are 
drawn in \sect{sec:conclusions}. 
\fi 
 
\section{Method} 
\label{sec:method} 
 
Predictions presented in this paper were obtained using events 
generated by \powhel~\cite{Garzelli:2011iu}, and stored in Les Houches 
event (LHE) files \cite{Alwall:2006yp}.  Our computational framework, 
\powhel, is an event generator that uses the \helacnlo\ codes
\cite{Bevilacqua:2011xh} (\helaconeloop\
\cite{Bevilacqua:2010mx} and \helacdipole\ \cite{Czakon:2009ss}) for
the computation of the amplitudes required as input by the \powhegbox\
\cite{Alioli:2010xd}. We use the latter for matching NLO QCD
computations with PS, according to the POWHEG method, generating events
stored in the Les Houches format (LHE).  Details of event generation
for \ttbb\ final states were presented in Ref.~\cite{Kardos:2013vxa}, where thorough checks of the correctness of the event files were also 
discussed. In particular, we treated the b-quarks massless, and used 
five massless flavours, but neglected the small contribution to the 
cross section from the b-quarks in the initial state. Feynman graphs 
containing $g\to \bB$ splittings are included in our computations. 
Such splittings are singular when the massless b-quarks become
collinear. To avoid this divergence, we employ a generation cut on the
invariant mass of the \bB\ pair, which is chosen such that it is
harmless when physical cuts are also employed \cite{Kardos:2013vxa}. 
We set the mass of the t-quark to $m_t = 173.2$\,GeV for the events at
8 TeV and to $m_t = 172.6$\,GeV for those at 14 TeV. The latter choice
was made in the first calculation at 14 TeV~\cite{Bevilacqua:2009zn},
against which we checked our predictions at NLO accuracy in 
Ref.~\cite{Kardos:2013vxa}.
 
Subsequent radiation emissions on top of the LHEs are generated by 
shower Monte Carlo (SMC) programs. SMCs can be further used to 
generate hadronization and hadron decay, as well as the decay of the 
heavy elementary particles in the narrow width approximation (NWA).  
For making predictions in this paper, we used the \pythia\ SMC 
code~\cite{Sjostrand:2006za}, in the Fortran version 6.4.28.  However, 
other SMCs can be used to further shower the events generated by
\powhel. 
 
Following Ref.~\cite{Kardos:2013vxa}, we fix the default value of the 
renormalization and factorization scales equal to $\mu_R=\mu_F=\mu_0=H_T/2$,
where $H_T$ is the sum of the transverse masses of partons in the final
state, using the underlying Born kinematics. Then we consider their
simultaneous variation by a factor of two around this value, leading to
scale uncertainty bands corresponding to scales in the range
$[\mu_0/2, 2\mu_0]$. We also consider $\mu_R$ and $\mu_F$ variations 
separately in this range and we verify that the uncertainties 
associated to off diagonal variations ($\mu_R, \mu_F/2$), ($\mu_R/2, \mu_F$),
($\mu_R, 2\mu_F$), ($2\mu_R, \mu_F$) lie 
within the bands of diagonal ones. Our scale uncertainty bands 
represent the envelope coming from these variations, leaving out the
antipodal choices ($\mu_R/2, 2\mu_F$) and ($2\mu_F, \mu_R/2$) according to
the prescription of Ref.~\cite{Cacciari:2012ny}. 
  
We used the {\texttt{CT10NLO}}, {\texttt{MSTW2008NLO}} and 
{\texttt{NNPDF}} sets of PDFs with 2-loop running $\alpha_S$ and 5 active flavours, as available in the {\texttt{LHAPDF}} interface~\cite{Whalley:2005nh}, 
drawing a PDF uncertainty 
band as the envelope of the results corresponding to the three central 
values of these PDF sets. For practical reasons, due to the intensive CPU 
requirements of our calculations, we neglected uncertainties coming from 
considering the statistical variations around each central PDF 
distribution. We expect that differences between different PDF sets 
are more pronounced than differences within each PDF set, as already 
found for the hadroproduction of \ttH\ final states in the HXSWG  
reports~\cite{Heinemeyer:2013tqa,Dittmaier:2011ti}.  
 
As discussed in Ref.~\cite{Kardos:2013vxa}, suppression factors are
used in order to suppress the generation of the events in those regions
of the phase space that are expected to be less relevant from the experimental
point of view, considering the typical cuts applied in the experimental
analyses. These factors however, lead to events characterized by a wide
weight distribution, extending without any explicit limit both in the
negative and in the positive weight region.  This may lead to spikes in
the kinematic distributions.  We expect these spikes would disappear
with significantly more statistics, but the present capabilities of
computer resources limit the total number of events we can generate to 
several millions. We thus implemented an automated spike-elimination 
procedure applied after SMC, relying on the fact that the same LHE can 
lead to different SMC emissions, i.e.~can populate different bins of 
the final differential distributions at the hadron level, depending on 
the random number sequence in the SMC generator.  
    
\section{Phenomenology} 
\label{sec:phenomenology} 
 
Using \powhel\ one can make predictions at four different stages in 
the evolution of the final state: (i) at the parton level using NLO 
accuracy, (ii) from the pre-showered POWHEG simulation (referred to 
Les Houches events, or LHEs), formally at the NLO accuracy, (iii) after 
decay of the heavy particles, (iv) at the hadron level after full SMC. 
A possible alternative to the last stage is (iv') after PS switching 
off hadronization and hadron decay, but in this 
case the heavy quarks are automatically kept stable by PYTHIA.  
Utilizing these options, we study the effect of these various stages  
of event evolution on several differential distributions,
before making predictions at the hadron level. 
 
\subsection{SMC effects} 
 
To understand the effect of the PS and hadronization, we first 
performed a phenomenological analysis where the cuts are applied on the 
LHEs, whereas distributions are plotted after different stages of the 
evolution. The cuts applied on jets formed from the LHEs, reconstructed 
using the \kt\ algorithm with $R=0.4$, as 
implemented in {{\texttt FastJet 3.0.6}}~\cite{Cacciari:2011ma} are the following: 
we require at least one $b$- and one $\bar{b}$-jet with 
(a) $p_T >$  20 GeV, 
(b)  $|\eta| <$  2.5, and 
(c)  $m_{b\bar{b}} >$ 100 GeV. 
We do not apply cuts on $b$-jets emerging from the decay of top quarks, 
implemented at LO accuracy in the NWA in the SMC, which neglects spin 
correlations.   
 
In \figss{fig:ptb1}{fig:dRbb} we present the comparisons of 
distributions at different stages of event evolution. Each plot 
contains five predictions, corresponding to the list (i--iv') 
presented in the introduction of this section. 
\begin{figure*}[t!] 
\includegraphics[width=0.9\linewidth]{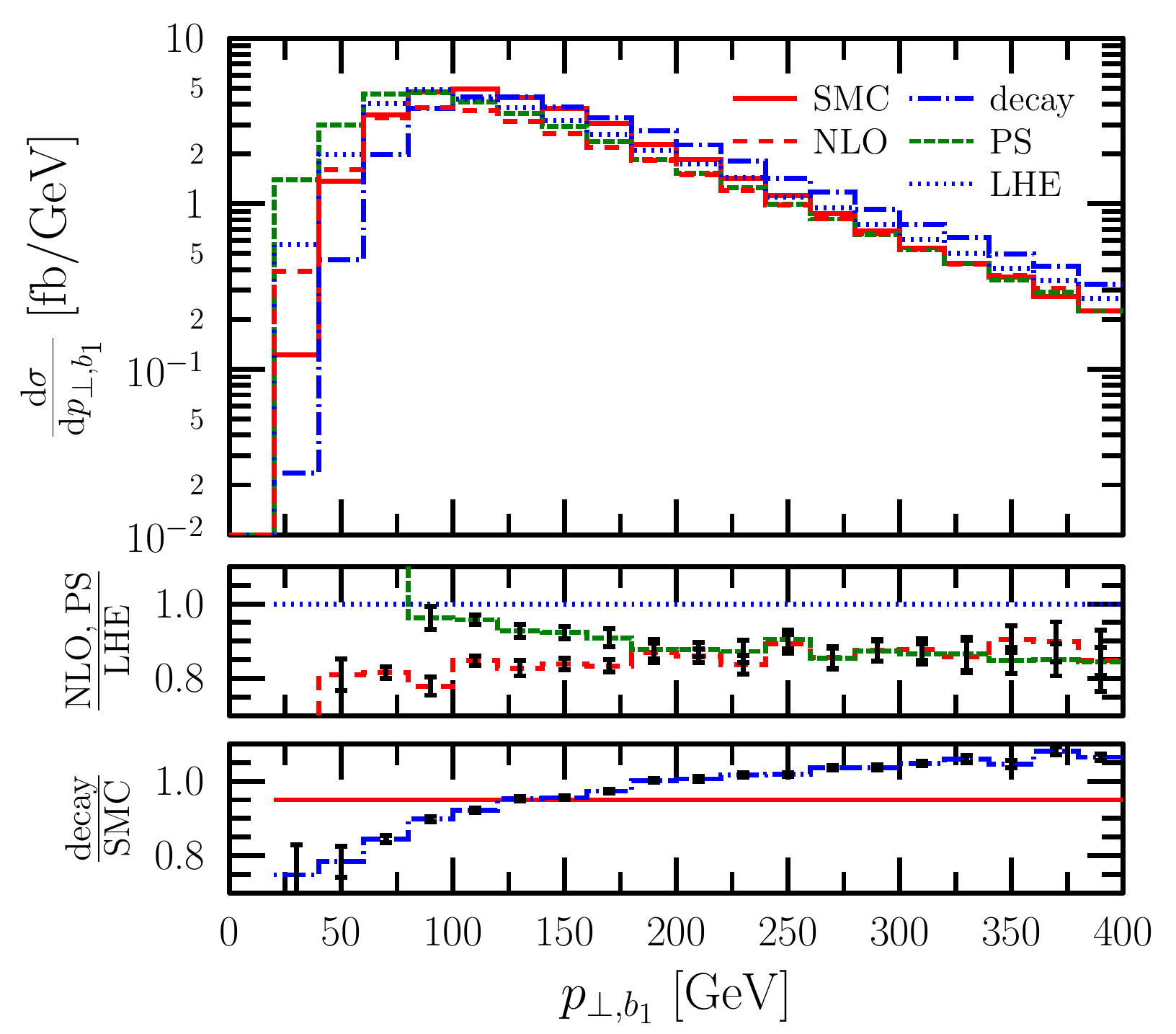} 
\caption{Distribution of transverse momentum of the 
hardest $b$-jet at the LHC at $\sqrt{s} = 14$\,TeV using 
\powhel. Distributions from LHEs are denoted LHE, while those at NLO 
accuracy by NLO. 
The middle panel shows the predictions at NLO accuracy, as well as after 
PS, normalized by the predictions from LHEs. 
The lower panel shows the predictions after decay normalized by the 
predictions after full SMC. 
The errorbars in both plots represent the combined statistical accuracy 
of the numerical integrations. 
} 
\label{fig:ptb1} 
\end{figure*} 
 
\begin{figure*}[t!] 
\includegraphics[width=0.9\linewidth]{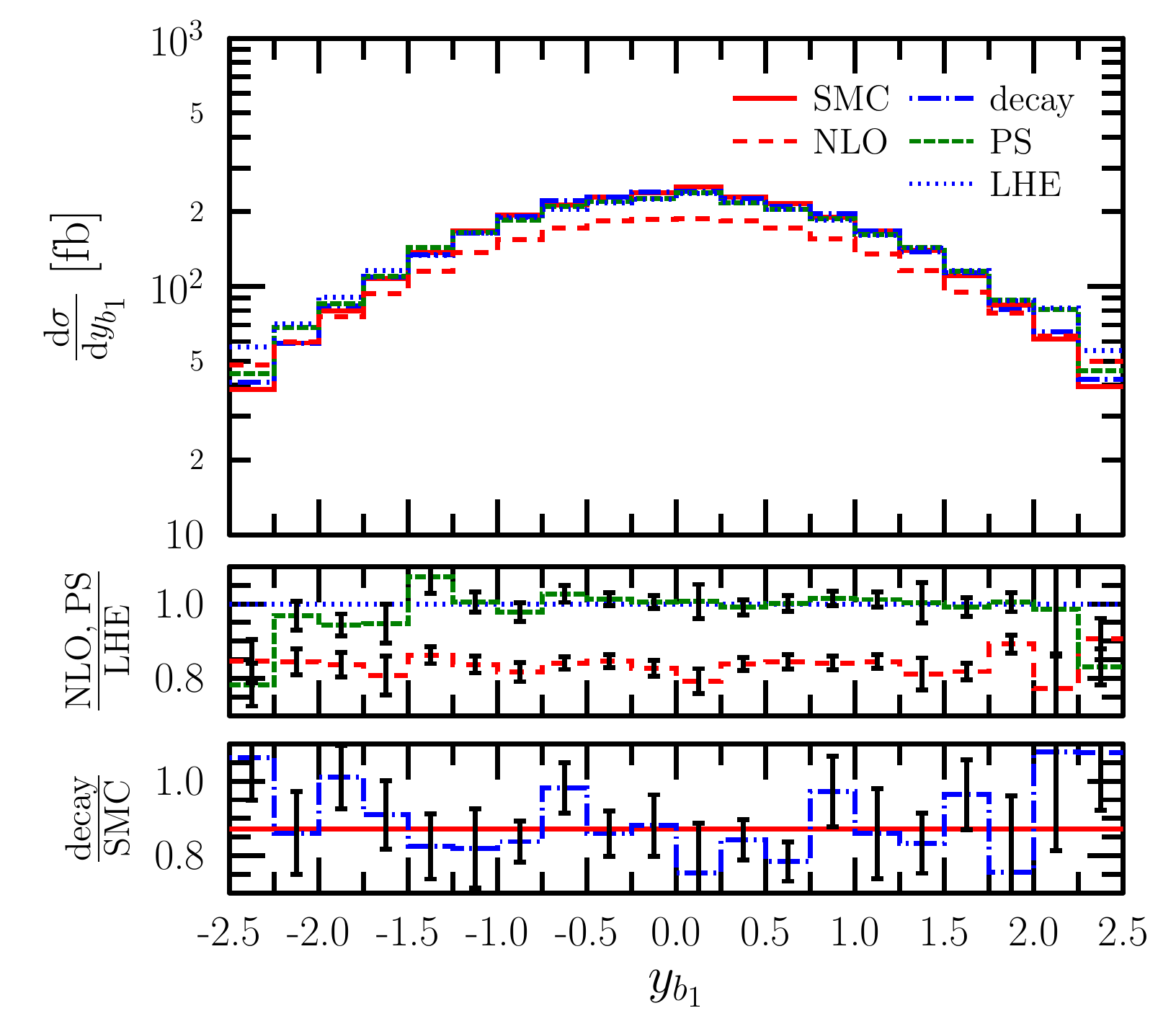} 
\caption{ 
Same as \fig{fig:ptb1}, as for the rapidity distribution of the hardest $b$-jet. 
} 
\label{fig:yb1} 
\end{figure*} 
 
\iffalse 
\begin{figure*}[t!] 
\includegraphics[width=0.49\linewidth]{SMCptb1.pdf} 
\hfill 
\includegraphics[width=0.49\linewidth]{SMCyb1.pdf} 
\caption{Distribution of (a) transverse momentum (b) rapidity of the 
hardest $b$-jet at the LHC at $\sqrt{s} = 14$\,TeV using 
\powhel. Distributions from LHEs are denoted LHE, while those at NLO 
accuracy by NLO. 
The middle panels show the predictions at NLO accuracy, as well as after 
PS, normalized by the predictions from LHEs. 
The lower panels show the predictions after decay normalized by the 
predictions after full SMC. 
The errorbars in both plots represent the combined statistical accuracy 
of the numerical integrations. 
} 
\label{fig:ptb1} 
\end{figure*} 
\fi 
 
\begin{figure*}[t!] 
\includegraphics[width=0.49\linewidth]{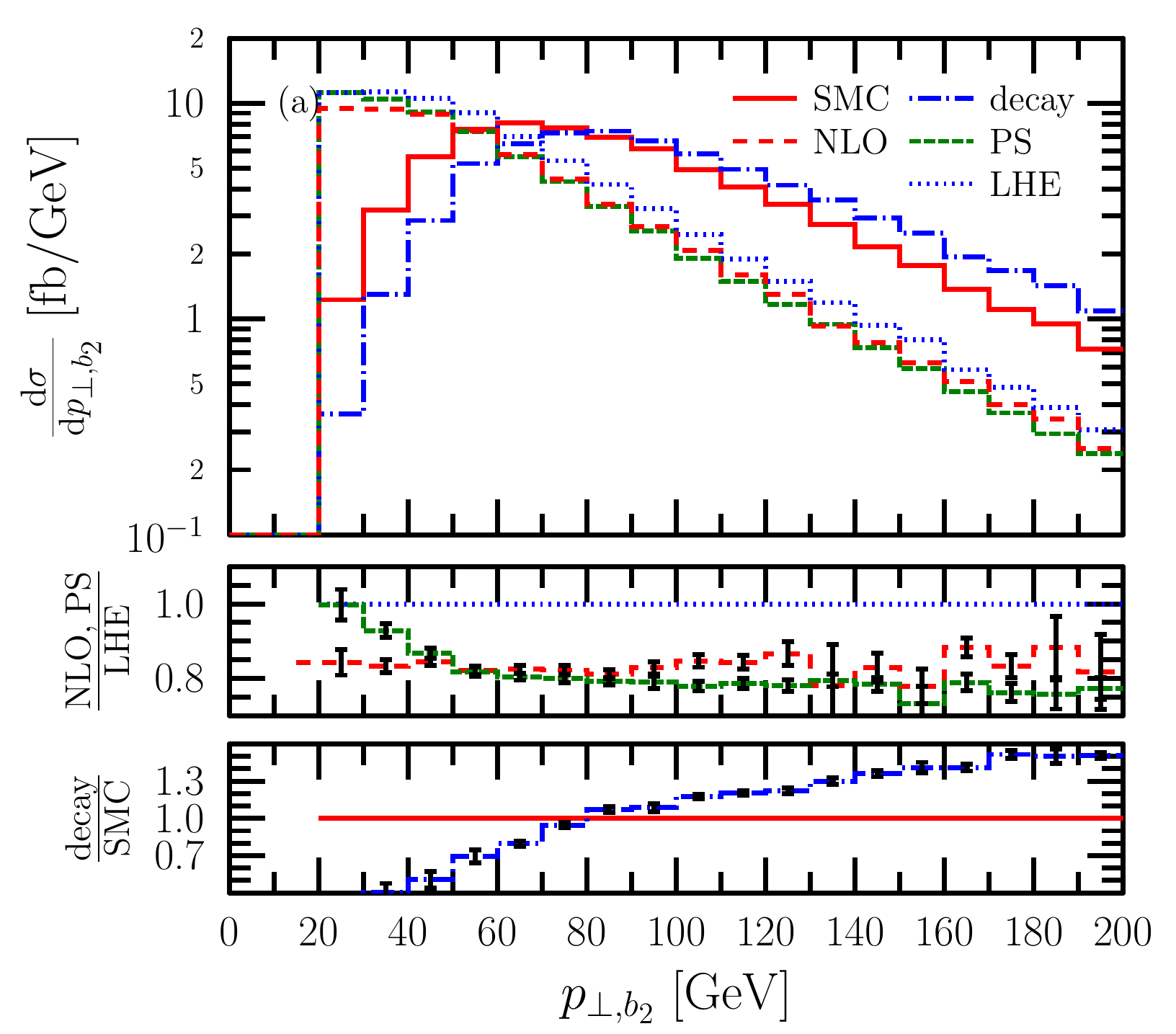} 
\hfill 
\includegraphics[width=0.49\linewidth]{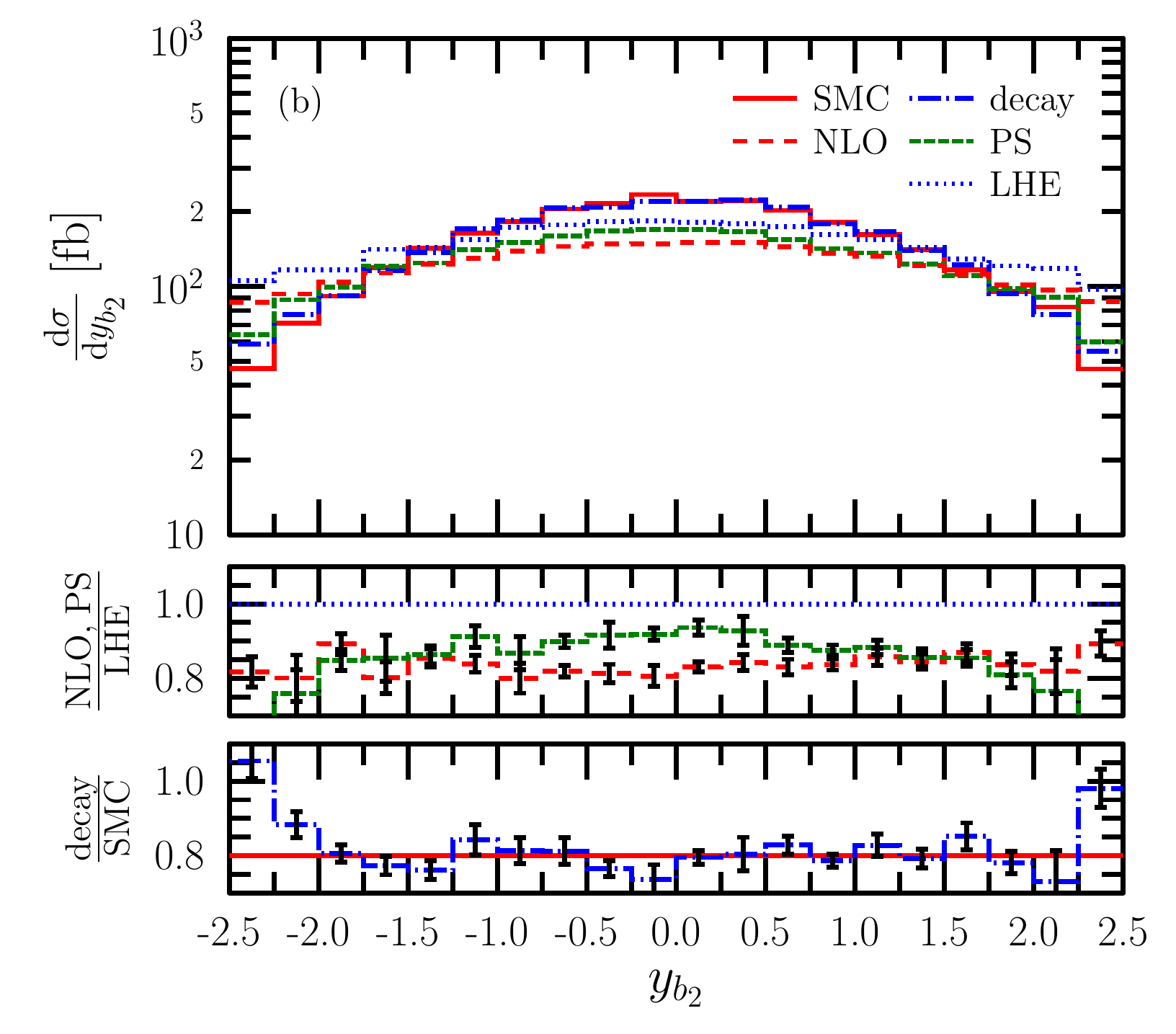} 
\caption{ 
Same as \figs{fig:ptb1}{fig:yb1}, as for the second hardest $b$-jet. 
} 
\label{fig:ptb2} 
\end{figure*} 
 
\begin{figure*}[t!] 
\centering 
\includegraphics[width=0.9\linewidth]{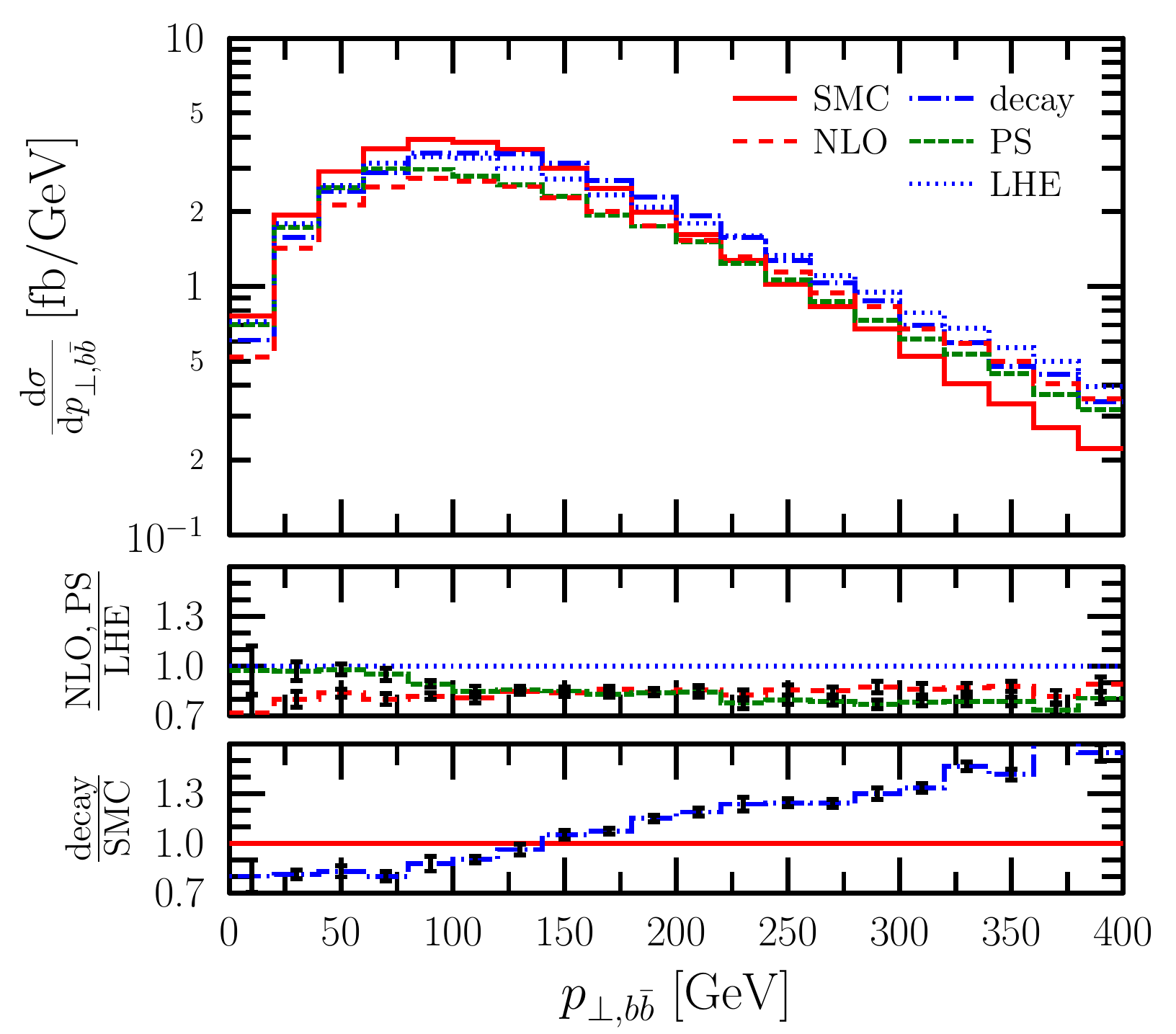} 
\caption{ 
Same as \fig{fig:ptb1}, as for the distribution of 
transverse momentum of the $b\bar{b}$-jet pair. 
} 
\label{fig:ptbb} 
\end{figure*} 
 
\begin{figure*}[t!] 
\centering 
\includegraphics[width=0.9\linewidth]{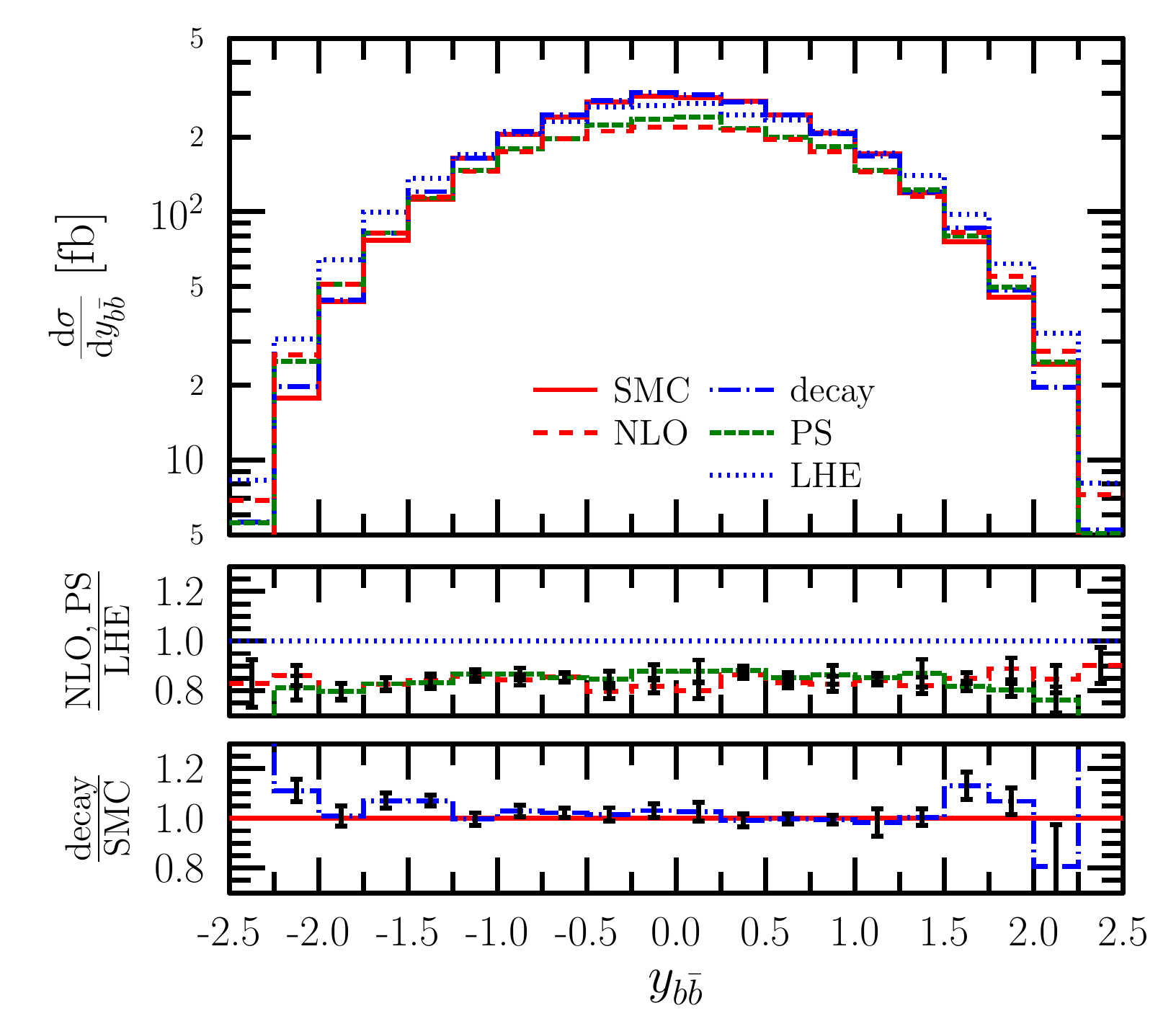} 
\caption{ 
Same as \fig{fig:ptb1}, as for the rapidity distribution of 
the hardest $b\bar{b}$-jet pair. 
} 
\label{fig:ybb} 
\end{figure*} 
 
\iffalse 
\begin{figure*}[t!] 
\includegraphics[width=0.49\linewidth]{SMCptbb.pdf} 
\hfill 
\includegraphics[width=0.49\linewidth]{SMCybb.pdf} 
\caption{ 
Same as \fig{fig:ptb1}, as for the distribution of 
(a) transverse momentum and (b) rapidity 
of the $b\bar{b}$-jet pair. 
} 
\label{fig:ptbb} 
\end{figure*} 
\fi 
 
\begin{figure*}[t!] 
\includegraphics[width=0.49\linewidth]{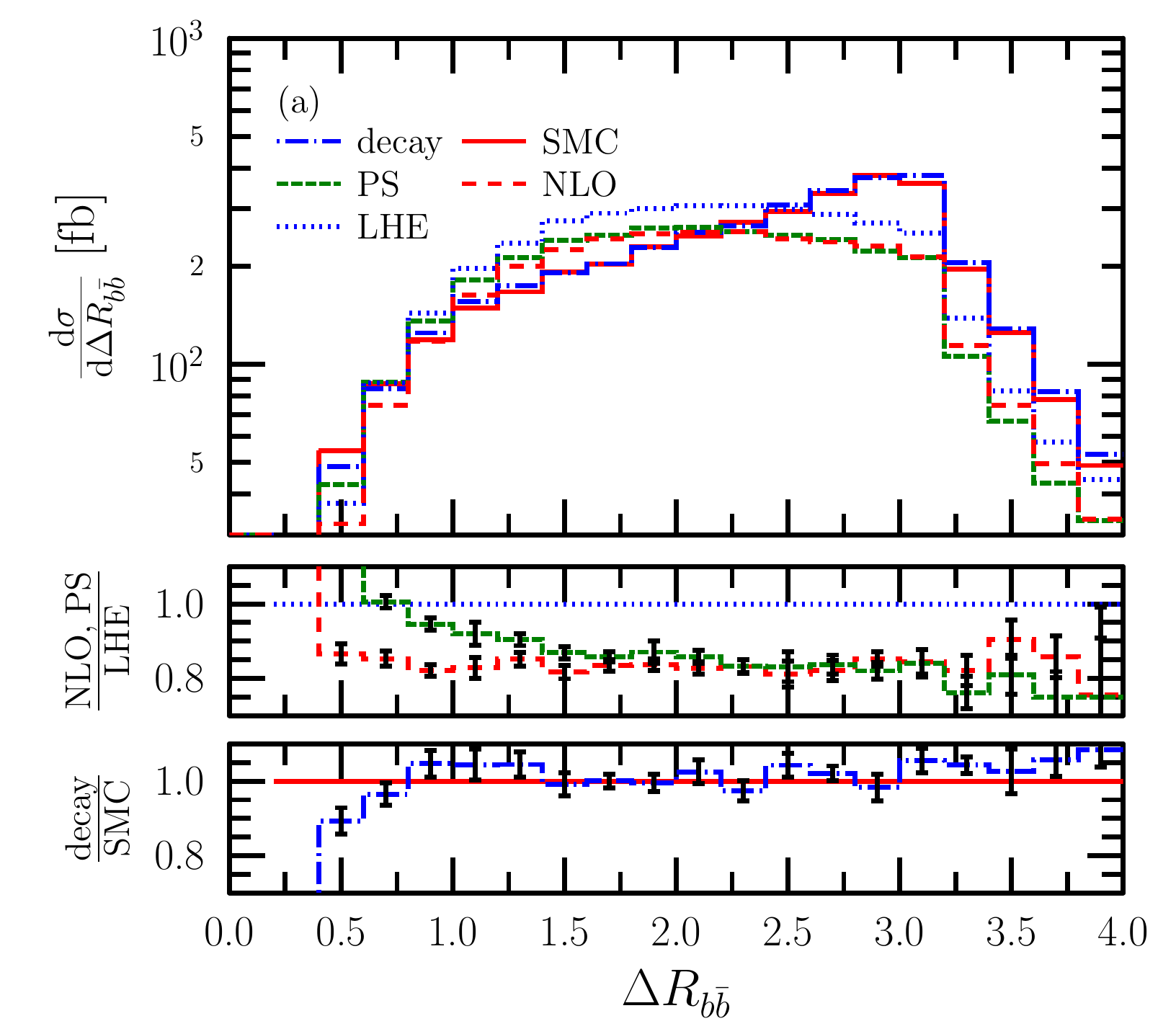} 
\hfill 
\includegraphics[width=0.49\linewidth]{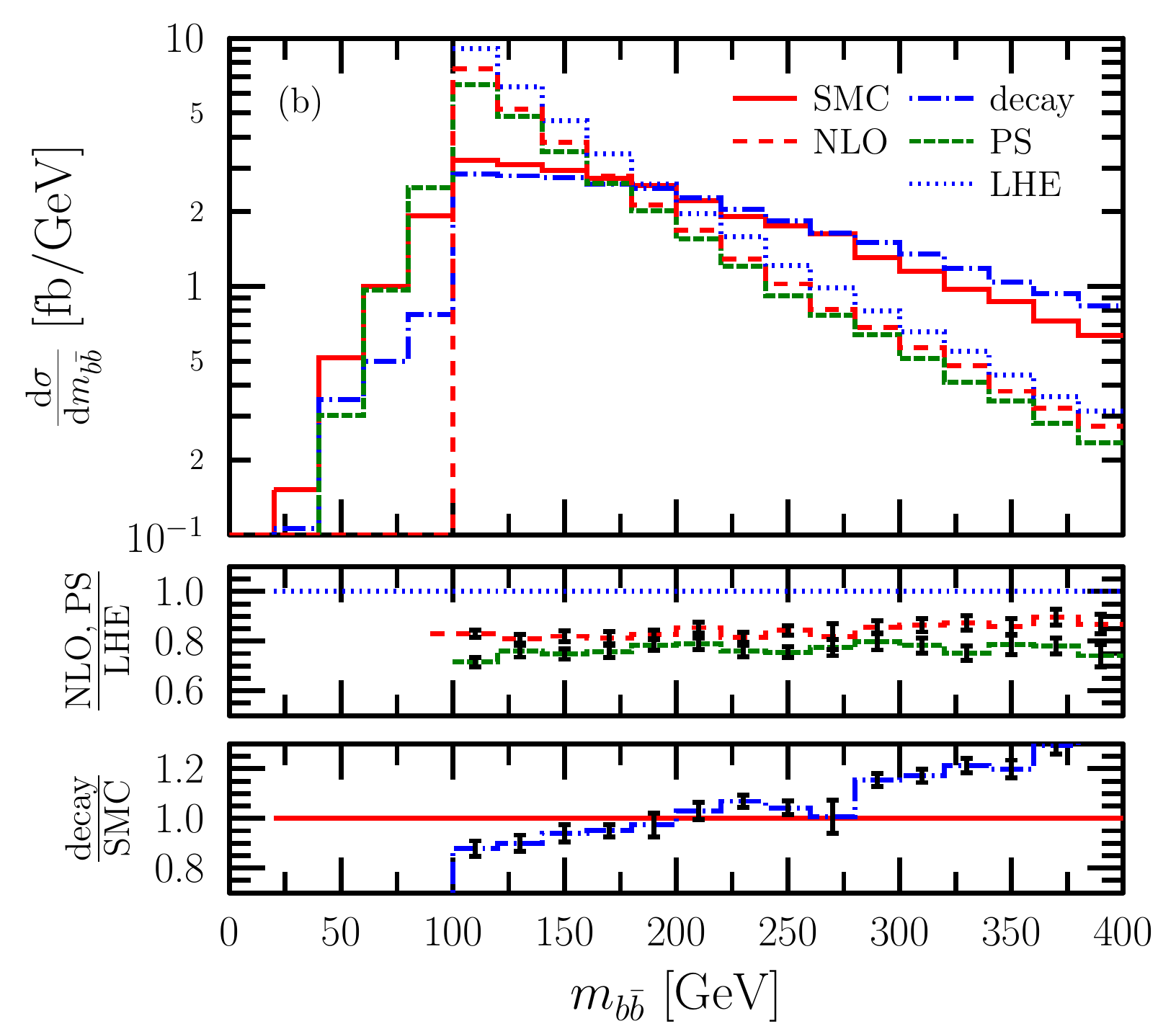} 
\caption{ 
Same as \fig{fig:ptb1}, as for the distribution of 
(a) $\Delta R_{b \bar{b}}$ separation and (b) invariant mass 
of the hardest $b\bar{b}$-jet pair. 
} 
\label{fig:dRbb} 
\end{figure*} 
 
First we study the effect of the PS by comparing differential 
distributions from the LHEs and after PS, keeping heavy particles 
stable. The corresponding predictions are marked as `LHE' and `PS'. In 
the middle section of each plot we also show the ratio PS/LHE of these 
predictions together with the ratio of the prediction at NLO accuracy 
to that from the LHEs, already studied in Ref.~\cite{Kardos:2013vxa}, 
where a fairly uniform increase of the distributions was found from the 
LHEs.  Here we find that the effect of the PS is to `bring back' 
the predictions close to the NLO ones with two exceptions: (i) the 
rapidity distribution of the hardest $b$-jet, which remains close to  
the prediction from LHEs even after shower evolution, and (ii) the 
distributions of transverse momenta for small $p_\bot$ (below about 
50\,GeV). Thus, the effect of the PS is in general small on top of the 
predictions at NLO level when the t-quarks are kept stable, or looking 
from an experimental point of view, when those are reconstructed from 
their decay products. 
 
Secondly we study the effect of the full SMC by comparing differential 
distributions from the LHEs, after decay and after SMC, corresponding 
to the full simulation including PS, hadronization and hadron decay. 
The corresponding predictions are labelled as `decay' and `SMC' and the 
ratio of these is also shown in the lower panel of each plot. Here we 
find significant changes in almost all distributions. The decay of the 
heavy quarks modifies the shapes, which are further modified by the SMC 
in the case of distributions of transverse momenta, or invariant mass 
of the $b\bar{b}$-jet pair. In case of rapidity distributions the 
effect of the SMC is negligible. 

\begin{figure}[t] 
\begin{center} 
\includegraphics[width=0.49\textwidth]{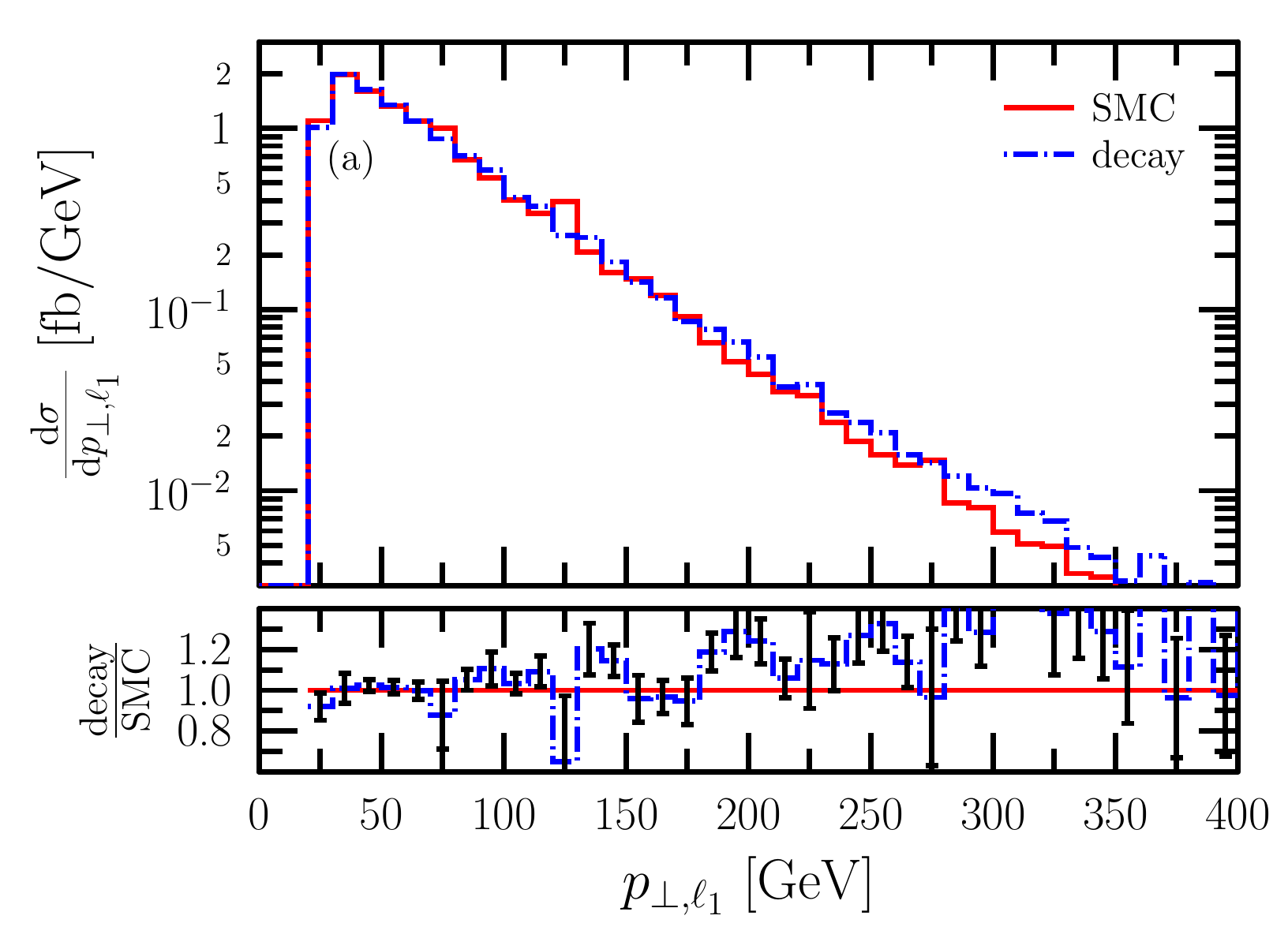} 
\includegraphics[width=0.49\textwidth]{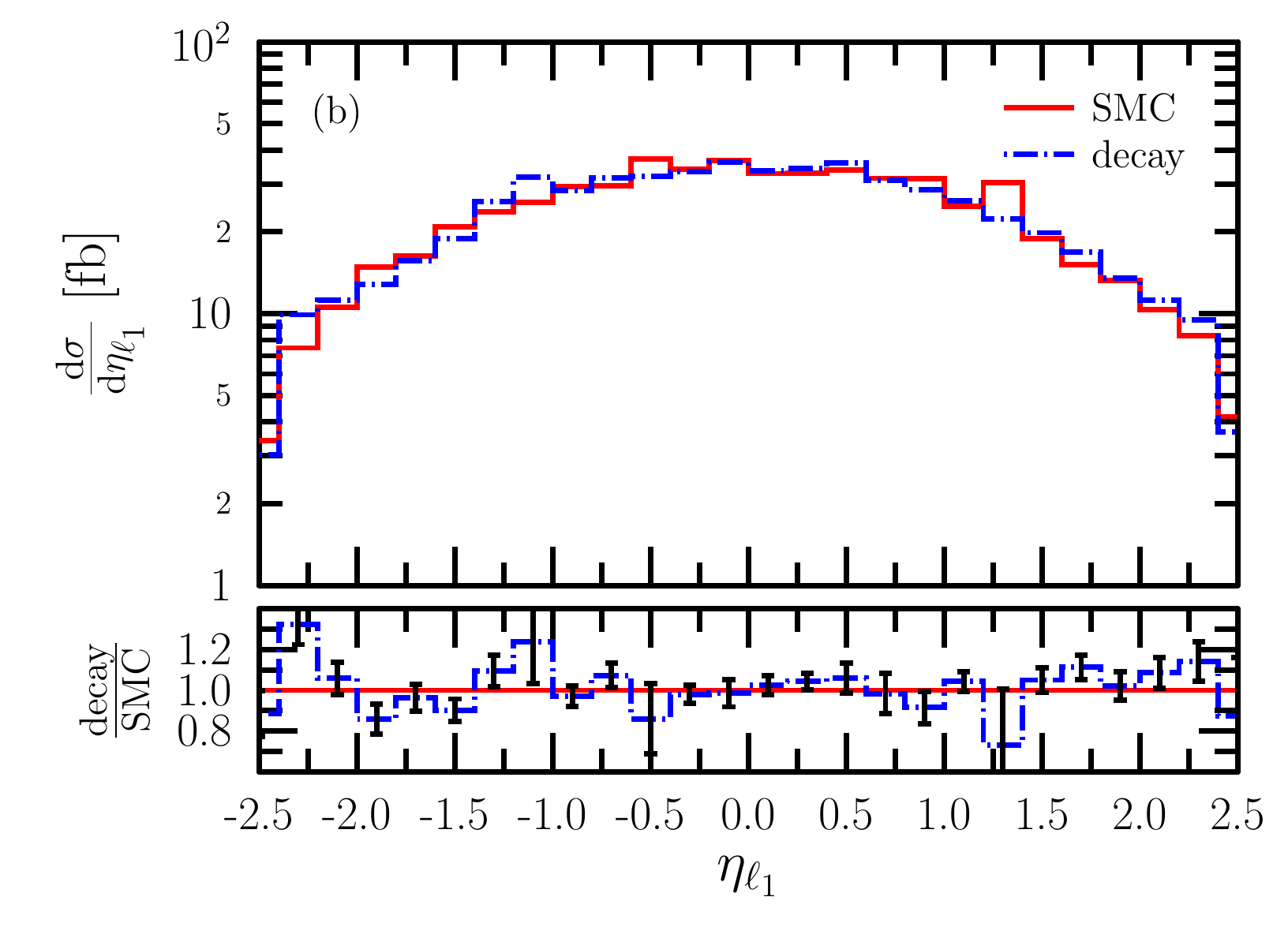} 
\end{center} 
\caption{\label{fig:lep} \powhel+\pythia\ predictions after decay and 
after SMC for (a) the transverse momentum and (b) the pseudorapidity of the 
hardest lepton. The lower panel shows the ratio of the predictions 
after SMC to those after decay.} 
\end{figure}

We select isolated leptons with $p_{\bot,\ell}>25$\,GeV, $|\eta_\ell|<2.5$
and minimal separation from all jets in the pseudorapidity-azimuthal
angle plane $\Delta R = 0.4$.  Looking at kinematic distributions of
those leptons, such as the transverse momentum and pseudorapidity of
the hardest isolated lepton plotted in 
Fig.~\ref{fig:lep}, it is clear that these are not affected by SMC 
effects, as expected. Such leptons appear in the decay of the heavy 
quarks, and neither the PS nor the hadronization change those. The same 
is true for the $W$ transverse mass, reconstructed for $W$'s decaying 
leptonically, not shown here.   
 
\begin{figure}[t] 
\begin{center} 
\centering 
\includegraphics[width=0.9\textwidth]{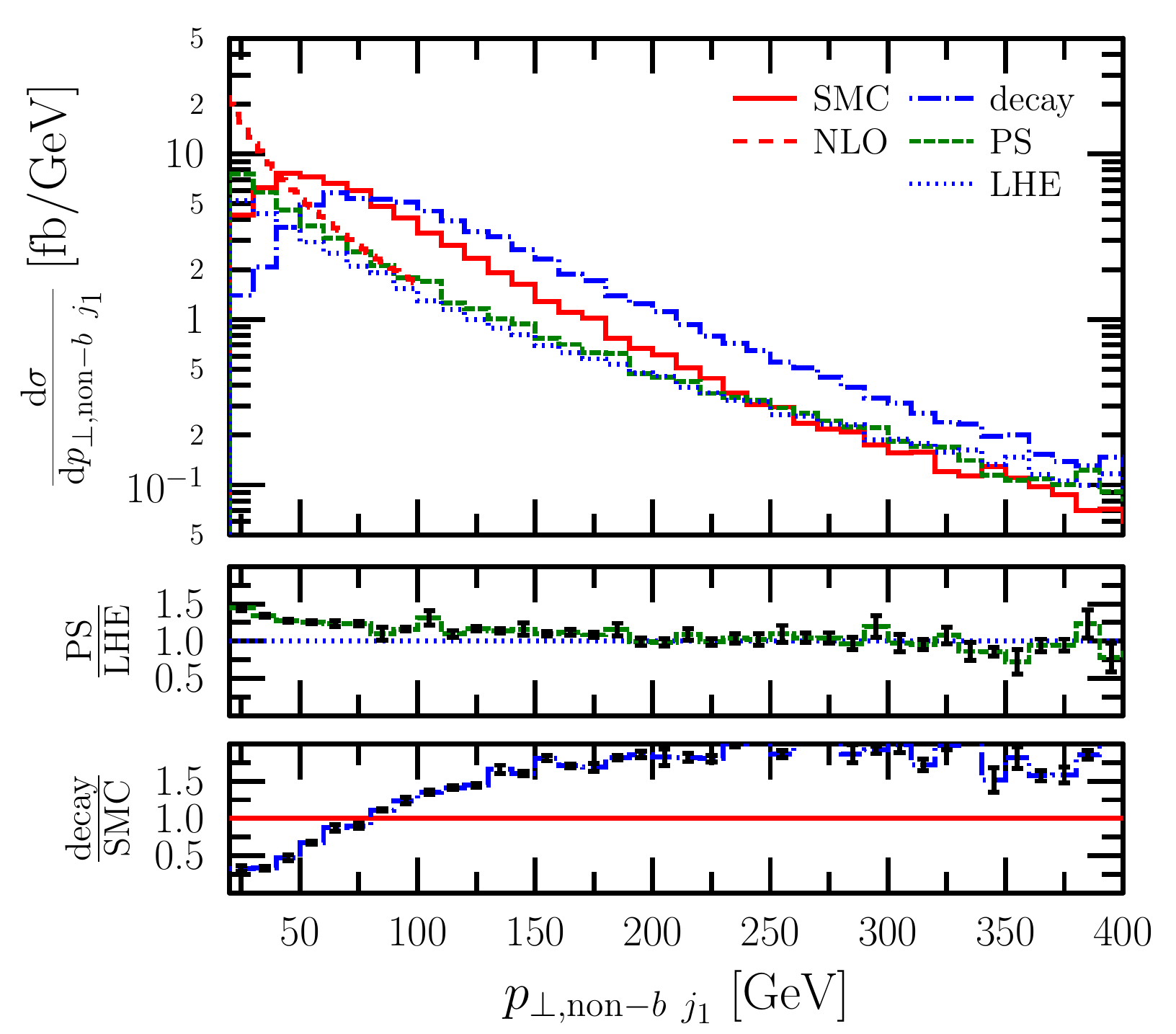} 
\end{center} 
\caption{\label{fig:extrajet} Same as \fig{fig:ptb1}, as for the
distribution of the transverse momentum of the hardest non-$b$ jet.  
} 
\end{figure}

On the other hand, the distributions most affected by the SMC are the 
total number of ($b$ + $\bar{b}$)-jets, the total number of jets ($b$ 
and non-$b$), the \pt\ of the hardest non-b jet. We show the last of 
these in Fig~\ref{fig:extrajet}. In this case the prediction at NLO 
accuracy diverges for small transverse momentum, which is screened by 
the Sudakov-factor, in case of parton shower matched prediction.  The 
decay of the heavy quarks hardens this distribution significantly 
through jets emerging from the hadronic decay of the $W$ bosons, which 
is subsequently softened by the PS and hadronization.   
\clearpage 
 
\subsection{Analyses with cuts at the hadron level} 
 
In order to make predictions at the hadron level including scale and 
PDF uncertainties, we decided to use selection cuts at the hadron 
level, inspired by the CMS note Ref.~\cite{CMS-PAS-TOP-13-010}, to
identify events in the dileptonic decay mode. These are rather
involved, which amply shows the flexibility of our method:  
\begin{enumerate} 
\item We require the events to have at least one pair of isolated 
opposite sign leptons with $p_{\bot,\ell}>20$\,GeV and $|\eta_\ell|<2.4$. 
Dilepton candidate events with invariant mass $m_{\ell\ell'} < 12$\,GeV 
are removed to suppress multijet final states, at essentially no 
penalty for the collected signal. A lepton is considered isolated if 
the sum of the transverse momenta of the charged hadrons, neutral 
hadrons and photons in a cone of $\Delta R < 0.3$ around it 
divided by its transverse momentum, $I_{\rm rel}$, 
does not exceed $I_{\rm rel}^{\max} = 0.15$.   
\item To remove the large background from events with $Z$-boson and 
jets, with $Z$ decaying into leptons, we require that the invariant 
mass of the lepton pair defined above ($e^+e^-$ or $\mu^+\mu^-$) 
falls outside a $\pm 15$\,GeV window centered at $m_Z$. 
\item Signal events typically contain large missing transverse energy 
due to the presence of neutrinos from decays of the $W$-bosons. We 
require a minimum missing transverse energy, $\pTmiss > 30$\,GeV for 
the $e^+e^-$ or $\mu^+\mu^-$ dilepton final states, but not for the 
mixed flavour dilepton final state. 
\item We cluster jets from hadrons, photons and non-isolated leptons
using the anti-\kt\ algorithm~\cite{Cacciari:2008gp} with $R=0.5$. 
The jets are all required to have $|\eta_j|<2.5$ and $p_{\bot,j}>40$ 
or 20\,GeV. We performed the analysis with both cuts, but present 
distributions here for the former only.
\item We require at least four well-separated $b$-jets after SMC, 
$\Delta R(j_i,j_k) > 0.5$ for $i,\,k \in$ \{1, 2, 3, 4\}, as well as 
$\Delta R(j_i,\ell_k) > 0.5$ for $i \in$ \{1, 2, 3, 4\}, and $k \in$ \{1, 2\} 
for the selected opposite-sign leptons. We assume  100\,\% b-tagging 
efficiency using {\tt MCTRUTH}. 
\end{enumerate} 
 
With these set of cuts the cross section for various choices of the  
default scale and for different PDF sets is presented 
in Tables~\ref{tab:CMS14pt20} and \ref{tab:CMS14pt40} for a collider
energy of 14\,TeV and in Table~\ref{tab:CMS8} for a collider energy of 8\,TeV.
These cross-section values were obtained at the hadron level by
showering events generated by \powhel\ with a $p_T$-ordered version of the 
\pythia\ PS, by using the Perugia 2011 tune~\cite{Skands:2010ak}. Using
the default version of \pythia, where the PS is virtuality-ordered in
absence of tunes, gives rise to values a few percent lower.  

In Table~\ref{tab:events} we exhibit the number of expected events for 
19.6\,fb$^{-1}$ at 8\,TeV. These marginally agree with the number of 
events expected in the CMS data sample, obtained on the basis of a LO+PS
computation by {\texttt{MadGraph}}~\cite{Alwall:2011uj} + \pythia\ (first
line in Table~2 of Ref.~\cite{CMS-PAS-TOP-13-010}). However, this
comparison can only be indicative for two reasons. On the one hand
our predictions assume 100\,\% b-tagging efficiency, while the
experimental analysis has smaller. On the other hand, we require at
least four $b$-jets, treating all $b$-jets on the same footing, while
the experimental analysis requires only two $b$-jets, because they do
not include the $b$-jets coming from t-quark decays in this count.
Furthermore, in order to identify $b$-jets we kept the lowest lying
B-hadrons stable in \pythia\ for simplicity, tagging as $b$-jets those
including at least a B-hadron, whereas the experiment reconstructs
$b$-jets from their decay products, using displaced vertex information. 

Recently an independent study of \ttbb\ and \ttjj\ hadroproduction, also
inspired by the same CMS note, appeared in Ref.~\cite{Bevilacqua:2014qfa}. This
work differs from our one because it is an NLO study, including a
simplified sets of cuts applied on parton level events (also including
top decays, but neglecting parton shower, hadronization and hadron decay
effects).
    
\begin{table}
\centering
\begin{tabular}{|c|c|c|c|c|c|}
\hline
\hline
$\mu_0$    & PDF        &   $e^+e^-$    &  $\mu^+\mu^-$   & $e^\pm\mu^\mp$ &      Total \bigstrut\\
\hline
\hline
$H_\bot$/4 & {\tt CT10} & 10.93$\pm$0.36 & 11.10$\pm$0.25 & 32.23$\pm$1.01 & 54.26$\pm$1.10 \bigstrut\\
\hline
$H_\bot$/2 & {\tt CT10} & 7.77$\pm$0.33 & 7.77$\pm$0.27 & 22.40$\pm$0.58 & 37.94$\pm$0.72 \bigstrut\\
\hline
$H_\bot$   & {\tt CT10} & 5.47$\pm$0.21 & 5.76$\pm$0.31 & 15.87$\pm$0.44 & 27.10$\pm$0.58 \bigstrut\\
\hline
$H_\bot$/2 & {\tt NNPDF}& 8.57$\pm$0.40 & 8.79$\pm$0.42 & 24.51$\pm$0.86 & 41.87$\pm$1.04 \bigstrut\\
\hline
$H_\bot$/2 & {\tt MSTW} & 9.04$\pm$0.45 & 8.65$\pm$0.24 & 24.48$\pm$0.86 & 42.17$\pm$1.00 \bigstrut\\
\hline
\hline
\end{tabular}
\caption{\label{tab:CMS14pt20} Total cross sections in fb at 14\,TeV in the
different dilepton channels with cuts listed in the text. The minimum
transverse momentum of the jets is set to 20\,GeV.~The quoted uncertainties are
of statistical origin.
}
\vspace{-0.1cm}
\end{table}
\begin{table} 
\centering 
\begin{tabular}{|c|c|c|c|c|c|} 
\hline 
\hline 
$\mu_0$    & PDF        &   $e^+e^-$    &  $\mu^+\mu^-$   & $e^\pm\mu^\mp$ &      Total \bigstrut\\ 
\hline             
\hline             
$H_\bot$/4 & {\tt CT10} & 2.71$\pm$0.04 & 2.72$\pm$0.04 & 7.81$\pm$0.06  & 13.24$\pm$0.09 \bigstrut\\ 
\hline             
$H_\bot$/2 & {\tt CT10} & 1.97$\pm$0.03 & 1.97$\pm$0.03 & 5.62$\pm$0.06  &  9.56$\pm$0.08 \bigstrut\\ 
\hline             
$H_\bot$   & {\tt CT10} & 1.43$\pm$0.03 & 1.42$\pm$0.02 & 4.19$\pm$0.10  &  7.04$\pm$0.11 \bigstrut\\ 
\hline             
$H_\bot$/2 & {\tt NNPDF}& 2.15$\pm$0.04 & 2.14$\pm$0.04 & 6.17$\pm$0.14  & 10.46$\pm$0.15 \bigstrut\\ 
\hline             
$H_\bot$/2 & {\tt MSTW} & 2.19$\pm$0.04 & 2.18$\pm$0.04 & 6.41$\pm$0.14  & 10.78$\pm$0.15 \bigstrut\\ 
\hline 
\hline 
\end{tabular} 
\caption{\label{tab:CMS14pt40} Total cross sections in fb at 14\,TeV in the 
different dilepton channels with cuts listed in the text. The minimum 
transverse momentum of the jets is set to 40\,GeV.} 
\end{table} 
\begin{table} 
\centering 
\begin{tabular}{|c|c|c|c|c|c|} 
\hline 
\hline 
$\mu_0$    & PDF        &   $e^+e^-$      &  $\mu^+\mu^-$     & $e^\pm\mu^\mp$  &     Total \bigstrut\\ 
\hline             
\hline             
$H_\bot$/4 & {\tt CT10} & 0.490$\pm$0.007 & 0.485$\pm$0.006 & 1.421$\pm$0.010 & 2.396$\pm$0.013 \bigstrut\\ 
\hline             
$H_\bot$/2 & {\tt CT10} & 0.346$\pm$0.006 & 0.343$\pm$0.004 & 1.002$\pm$0.006 & 1.691$\pm$0.010 \bigstrut\\ 
\hline             
$H_\bot$   & {\tt CT10} & 0.246$\pm$0.004 & 0.242$\pm$0.003 & 0.714$\pm$0.008 & 1.204$\pm$0.010 \bigstrut\\ 
\hline                         
$H_\bot$/2 & {\tt NNPDF}& 0.361$\pm$0.006 & 0.360$\pm$0.004 & 1.060$\pm$0.013 & 1.781$\pm$0.015 \bigstrut\\ 
\hline                         
$H_\bot$/2 & {\tt MSTW} & 0.374$\pm$0.005 & 0.376$\pm$0.004 & 1.098$\pm$0.017 & 1.848$\pm$0.018 \bigstrut\\ 
\hline 
\hline 
\end{tabular} 
\caption{\label{tab:CMS8} Total cross sections in fb at 8\,TeV in the 
different dilepton channels with cuts listed in the text. The minimum 
transverse momentum of the jets is set to 40\,GeV.} 
\end{table} 
\begin{table} 
\centering 
\begin{tabular}{|c|c|c|c|} 
\hline 
\hline 
  $e^+e^-$ &  $\mu^+\mu^-$ & $e^\pm\mu^\mp$  &     Total \bigstrut\\ 
\hline             
\hline             
\res{6.8}{+2.8}{-2} & \res{6.7}{+2.8}{-2} & \res{19.6}{+8.2}{-5.4}   & 
\res{33.1}{+13.8}{-9.4}  \bigstrut\\ 
\hline  
\hline 
\end{tabular} 
\caption{\label{tab:events} Number of expected \ttbb\ events at 8\,TeV 
in the different dilepton channels with cuts listed in the text. The 
minimum transverse momentum of the jets is set to 40\,GeV. The 
uncertainty corresponds to the variation of the renormalization and 
factorization scales around their default value by factors of half and two.} 
\end{table} 
 
\begin{figure*}[t!] 
\includegraphics[width=0.49\linewidth]{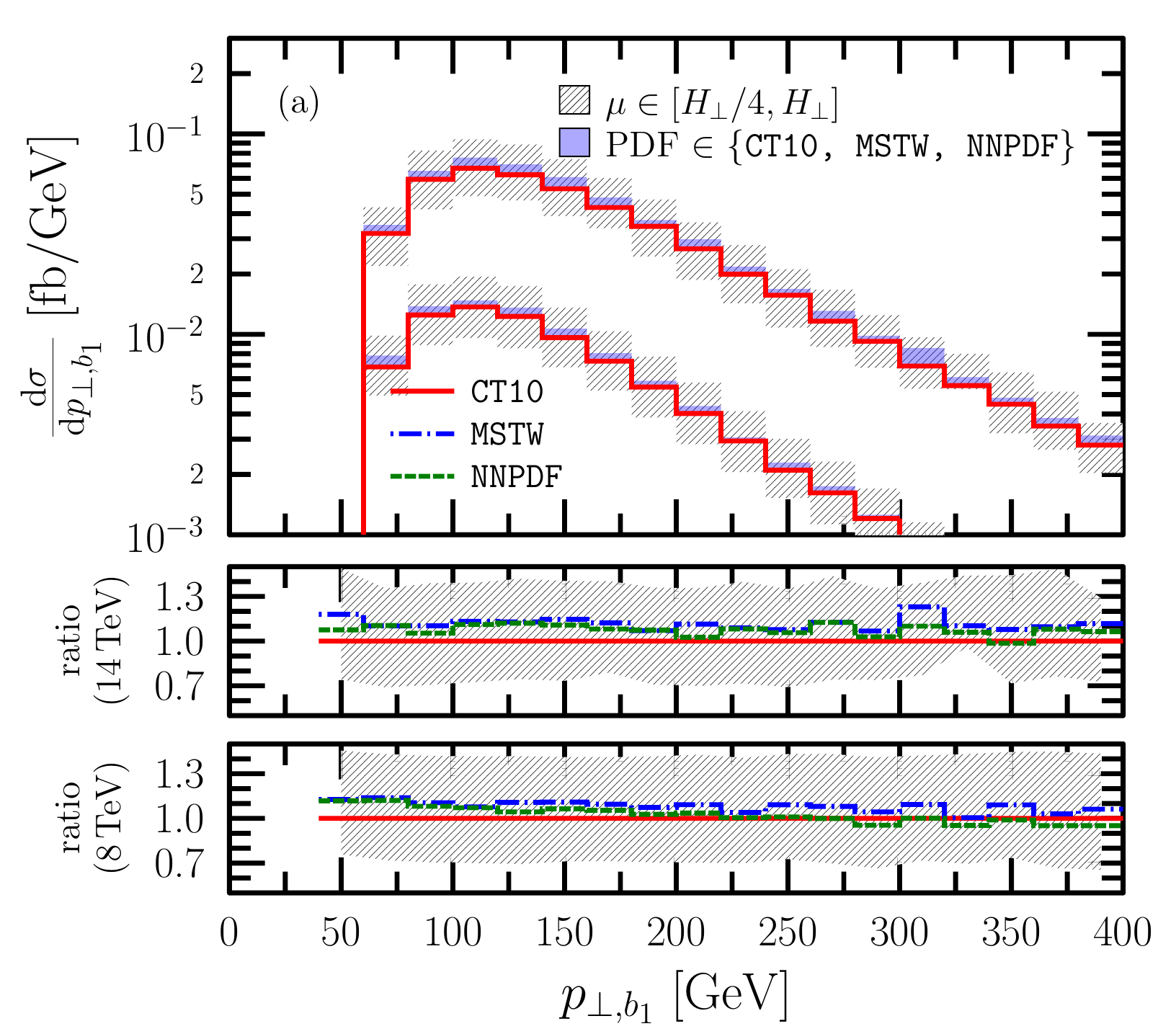} 
\hfill 
\includegraphics[width=0.49\linewidth]{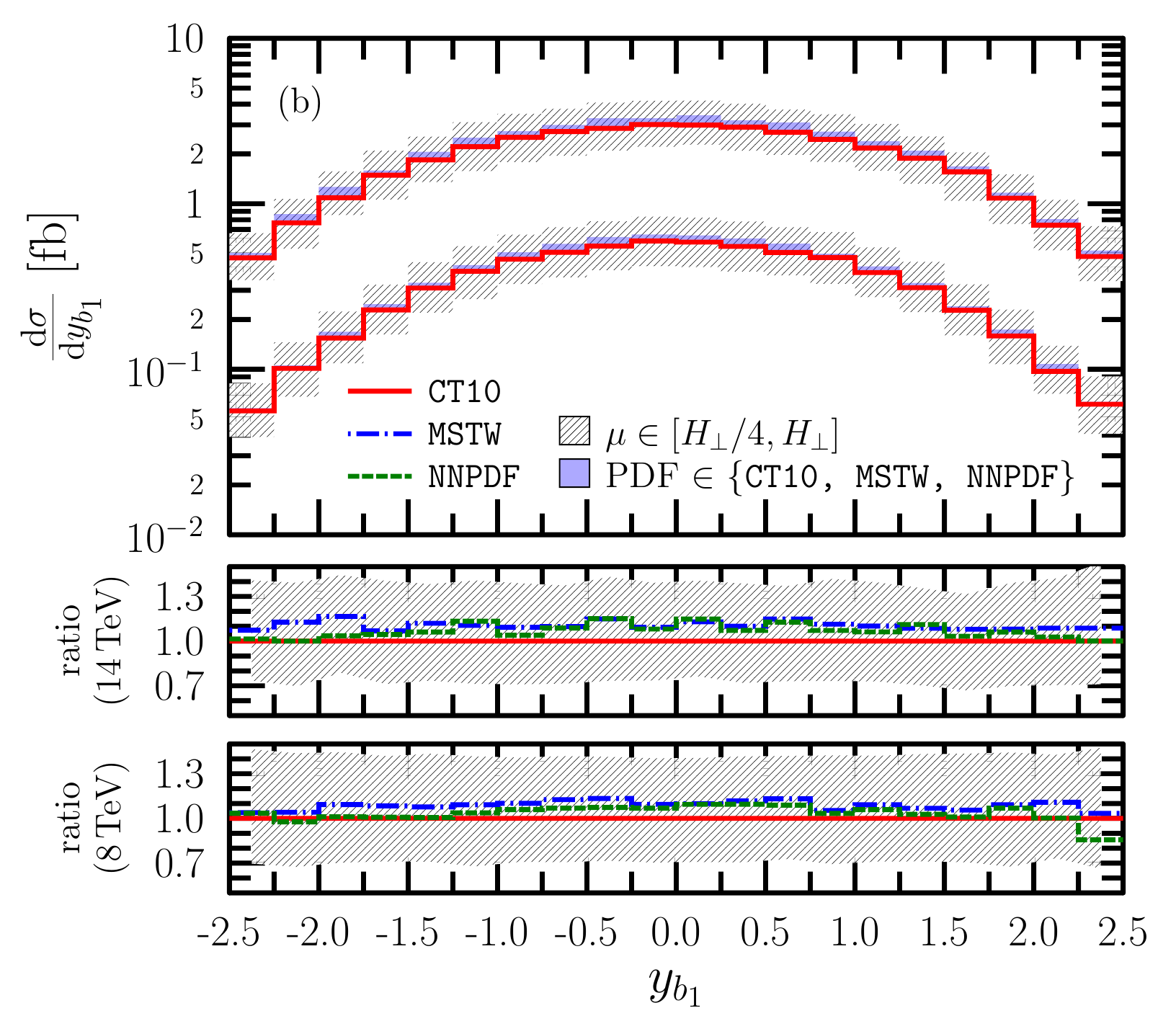} 
\caption{Distribution of (a) transverse momentum (b) rapidity of the 
hardest $b$-jet at the LHC at $\sqrt{s} = 14$\,TeV using \powhel\ + \pythia\ after full SMC, under the cuts (1 -- 5) listed in the text. The band represent the envelope of the scale uncertainties.  
The lower panels show the ratio of predictions obtained with {\texttt{MSTW2008NLO}} and {\texttt{NNPDF}} sets to that obtained with {\texttt{CT10NLO}}. 
} 
\label{fig:CMSptb1} 
\end{figure*}

In \figss{fig:CMSptb1}{fig:CMSmbb} we present several kinematic 
distributions obtained with these selection cuts: the transverse momenta 
and rapidities of the two hardest $b$-jets, the transverse momentum, 
rapidity and invariant mass distribution of the hardest $b$-jet pair and 
the invariant mass distribution of the hardest $b$-jet and hardest 
charged lepton. In each figure we show two predictions, one for 
8\,TeV c.m.~energy (smaller cross sections) and one for 14\,TeV 
(larger cross sections). The shapes of the rapidity distributions are 
very similar at the two energies, the result of the larger 
collision energy just amounting to about a nine-fold increase 
of the cross section. On the other hand, in case of the invariant mass and transverse momentum distributions, the spectra at 14\, TeV also become considerably harder with respect to those at 8\, TeV.   
 
\begin{figure*}[t!] 
\includegraphics[width=0.49\linewidth]{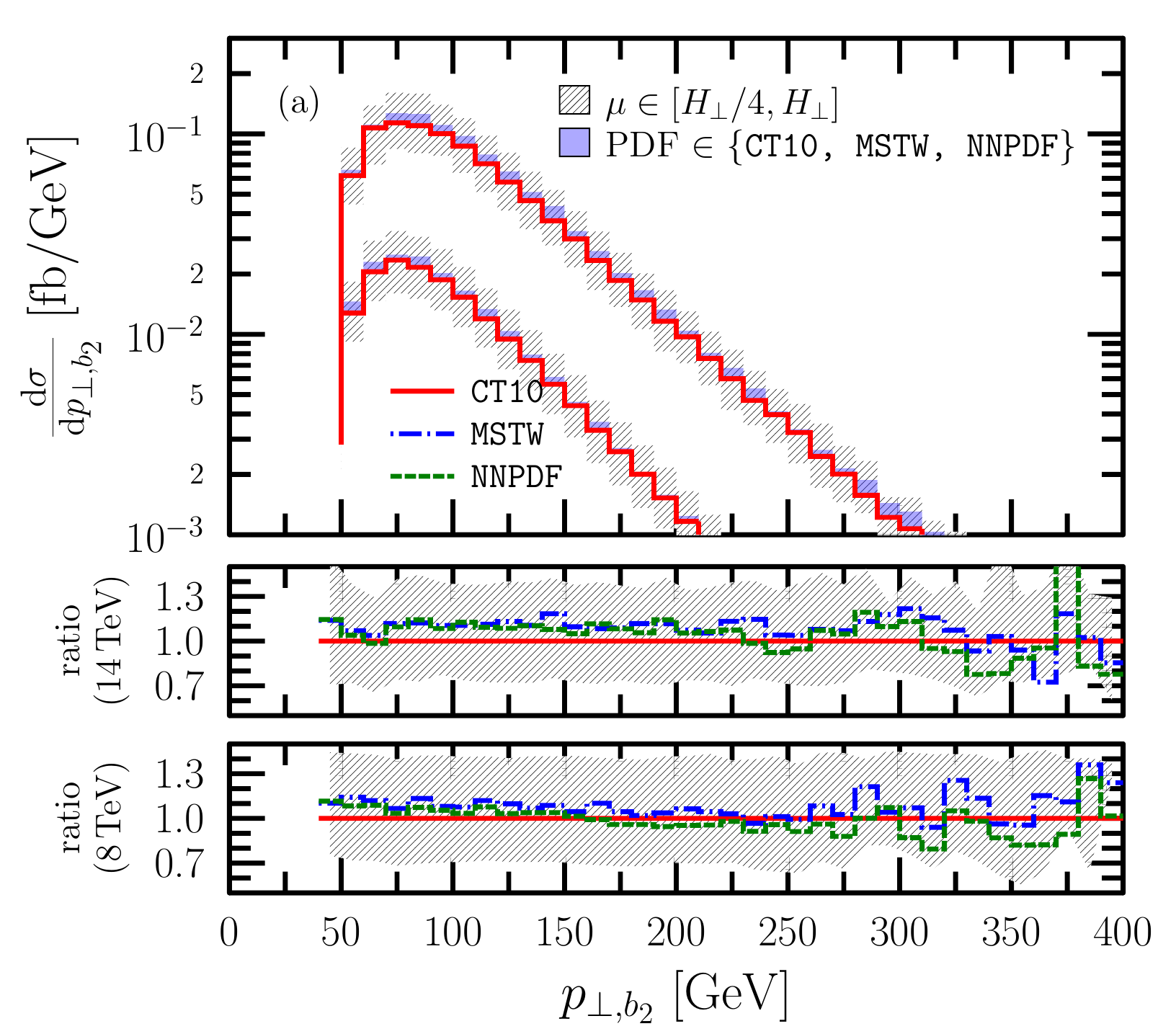} 
\hfill 
\includegraphics[width=0.49\linewidth]{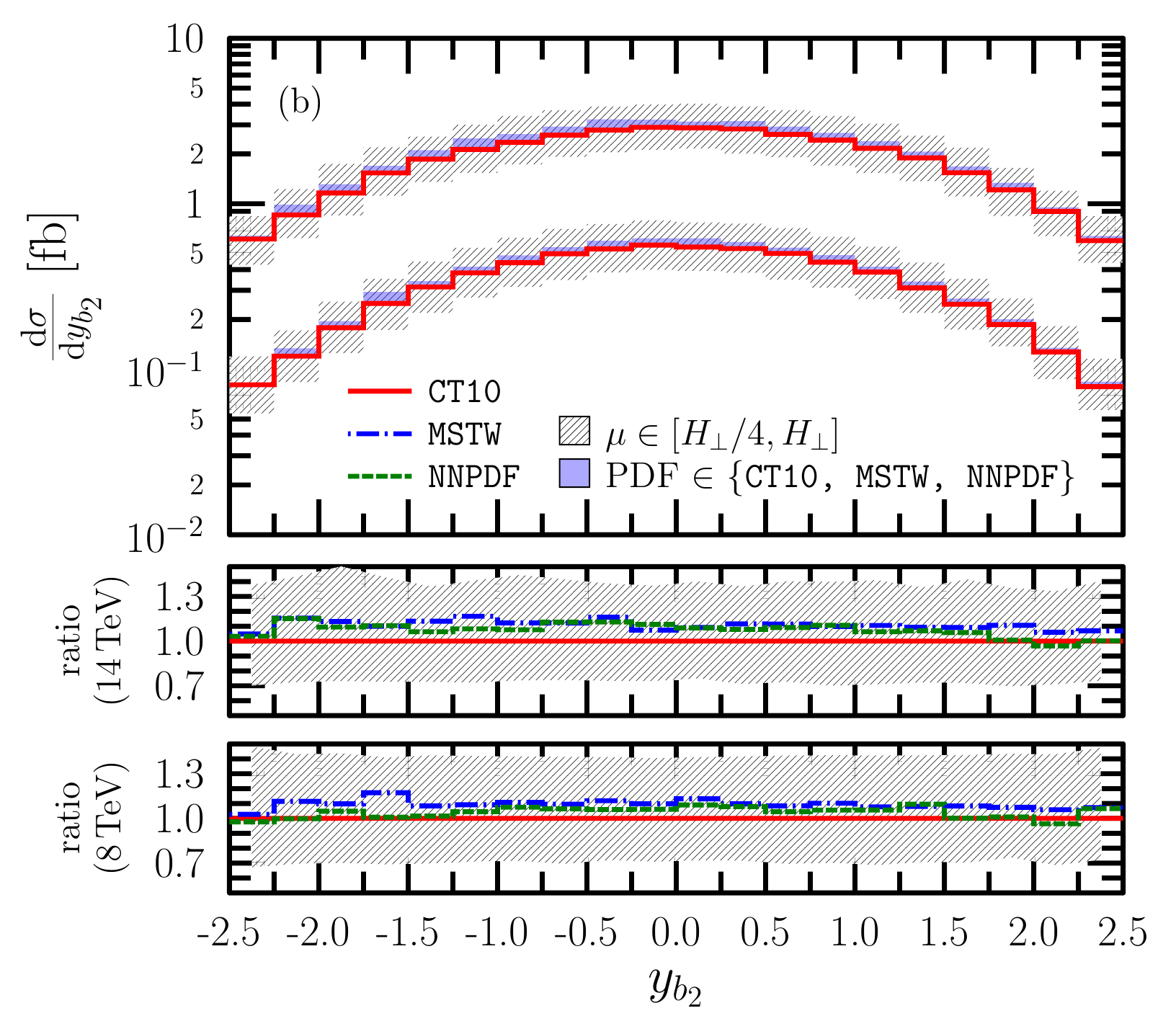} 
\caption{Same as \fig{fig:CMSptb1}, as for the second hardest $b$-jet. 
} 
\label{fig:CMSptb2} 
\end{figure*} 
 
\begin{figure*}[t!] 
\includegraphics[width=0.49\linewidth]{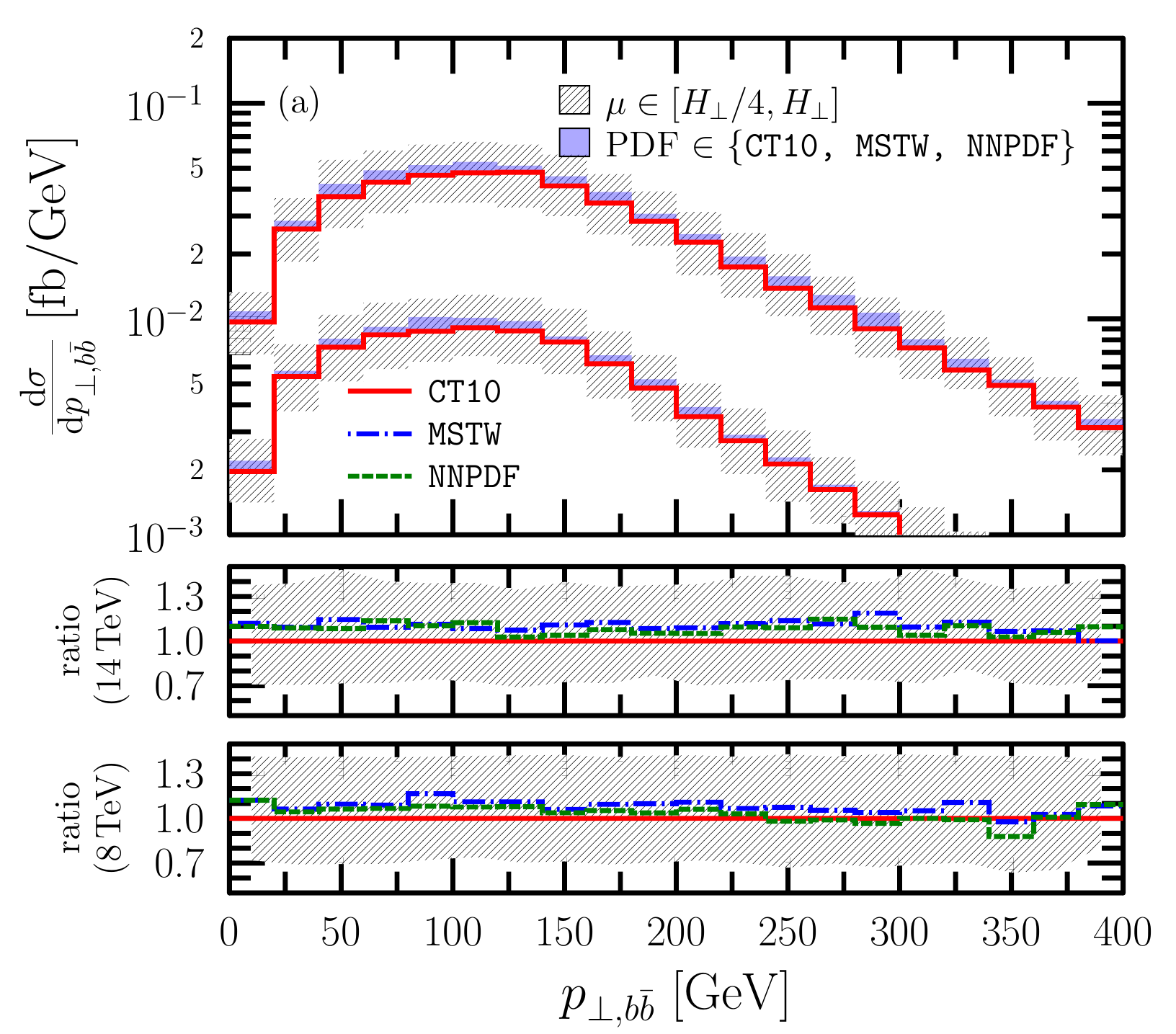} 
\hfill 
\includegraphics[width=0.49\linewidth]{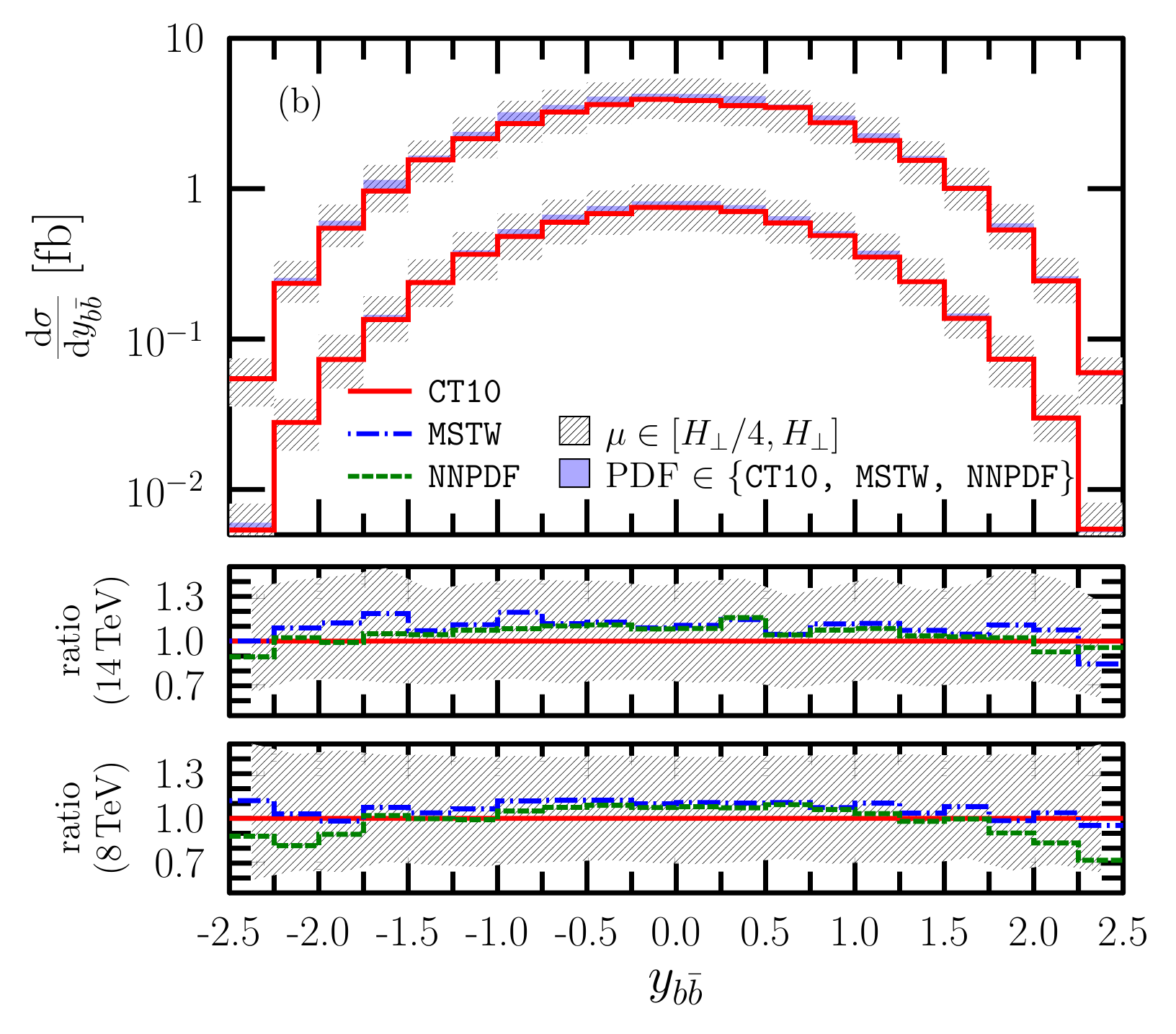} 
\caption{Same as \fig{fig:CMSptb1}, as for the hardest $b\bar{b}$-jet pair. 
} 
\label{fig:CMSptbb} 
\end{figure*} 
 
\begin{figure*}[t!] 
\includegraphics[width=0.49\linewidth]{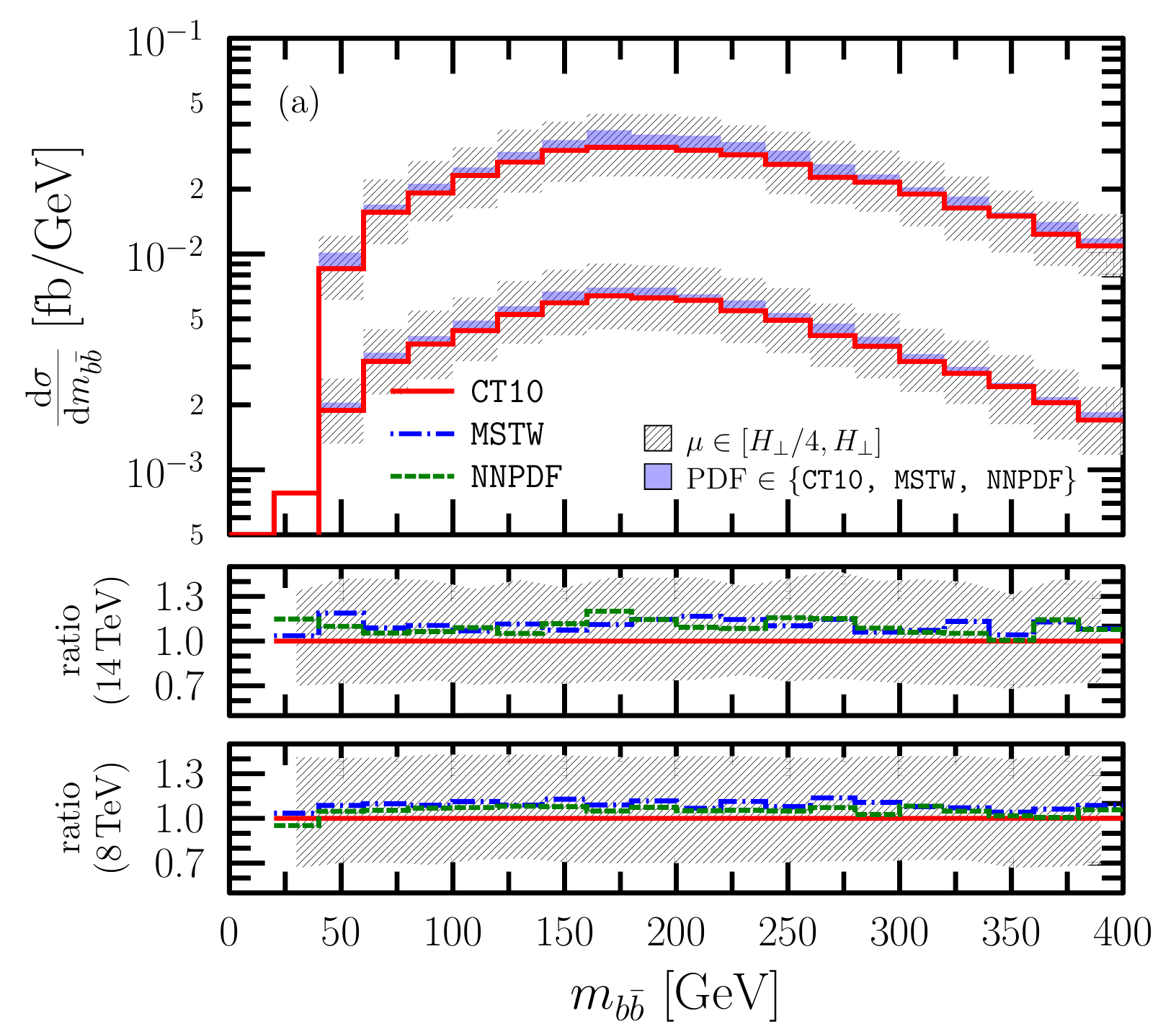} 
\hfill 
\includegraphics[width=0.49\linewidth]{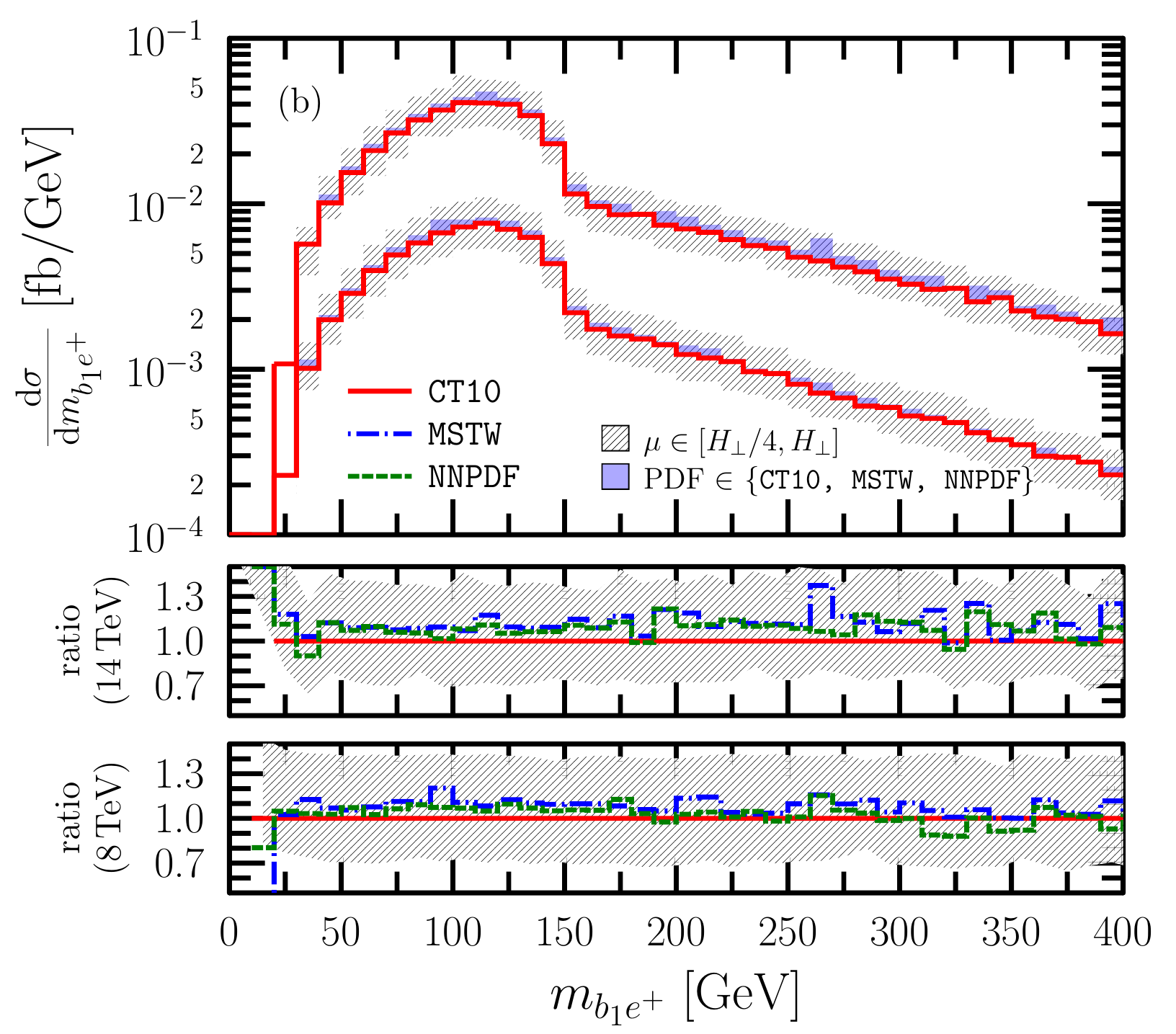} 
\caption{Same as \fig{fig:CMSptb1}, as for the invariant mass of  
(a) the hardest $b\bar{b}$-jet pair, (b) the hardest $b$-jet and positron.} 
\label{fig:CMSmbb} 
\end{figure*} 
 
\begin{figure*}[t!] 
\includegraphics[width=0.49\linewidth]{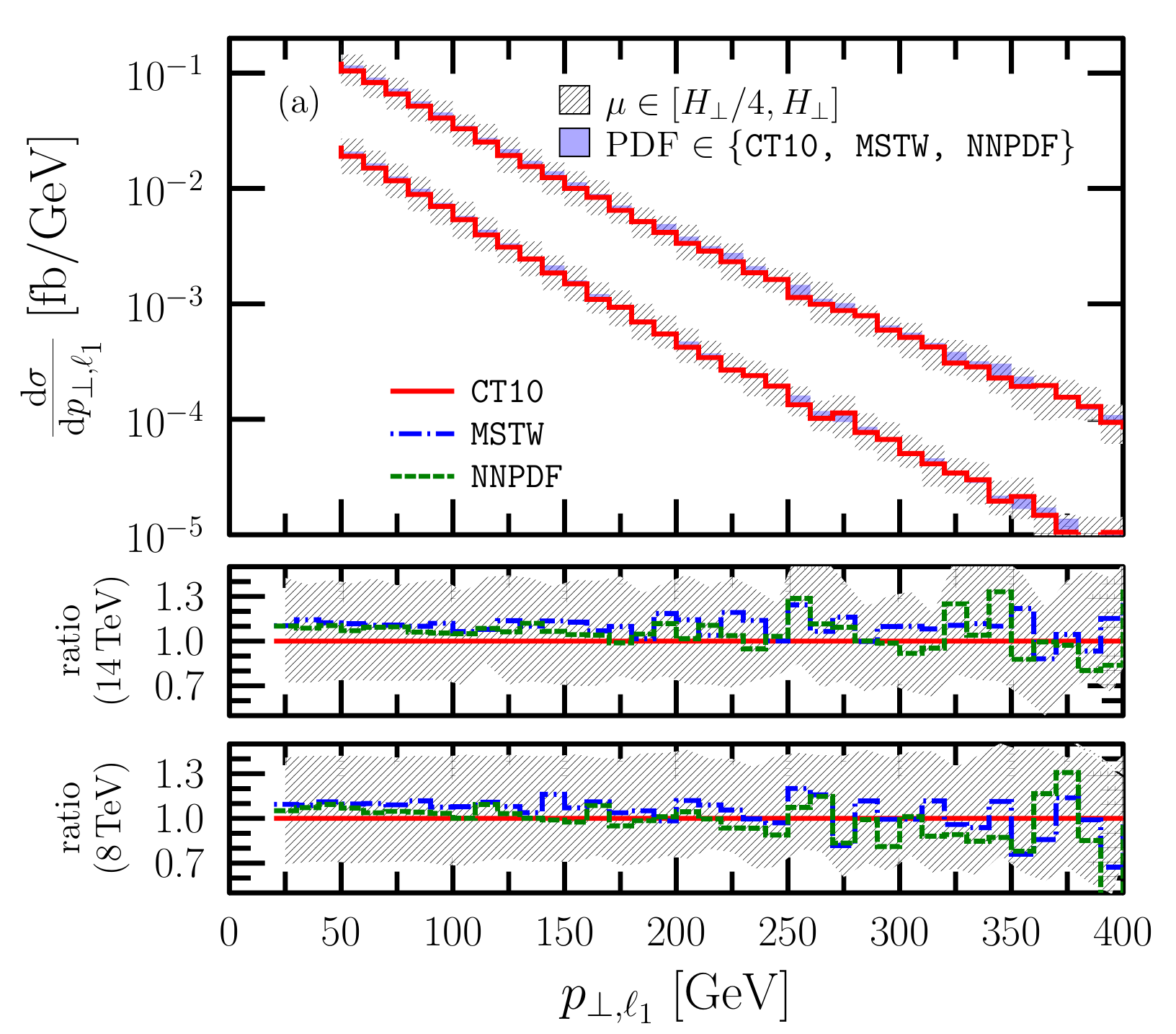} 
\hfill 
\includegraphics[width=0.49\linewidth]{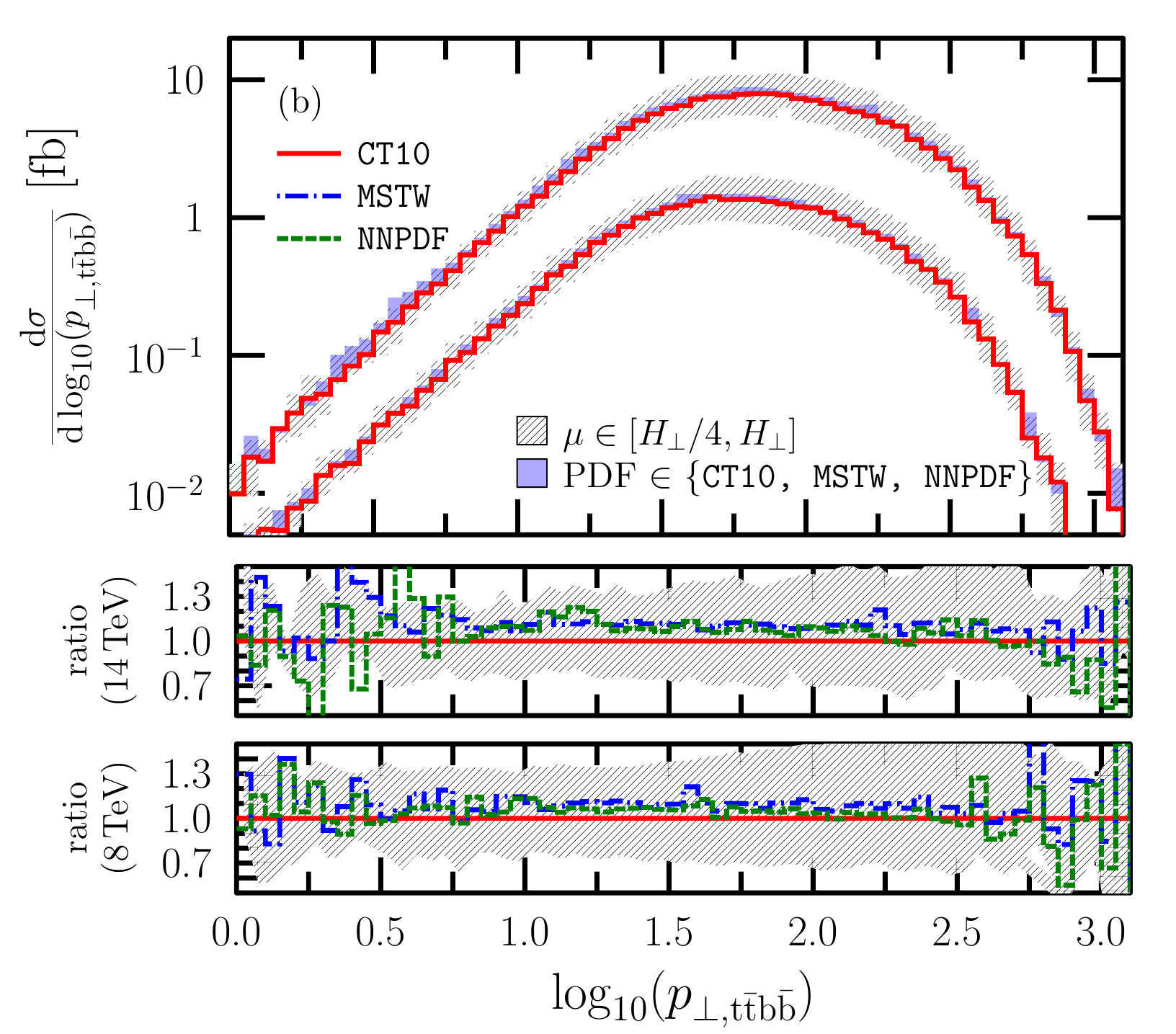} 
\caption{Same as \fig{fig:CMSptb1}, as for the transverse momentum of   
(a) the hardest lepton, (b) the \ttbb\ system.} 
\label{fig:CMSptl1} 
\end{figure*} 
 
In the same figures we exhibit two bands: one corresponding to the 
envelope of standard variations of the renormalization and 
factorization scales, between half and twice the default scale $\mu_0$, 
while the other to the envelope of dependence on the PDF set ({\tt CT10NLO}, 
{\tt MSTW2008NLO} and {\tt NNPDF}). In order to see these better, in the lower 
panels we show all predictions normalized to our default choice: 
predictions with $\mu_0 = H_\bot/2$ and {\tt CT10NLO} PDF set. The middle 
panels corresponds to the cross sections at 14\,TeV, while the lower ones 
to those at 8\,TeV. 
 
We see that the scale dependence for these distributions
follows the scale dependence found at 
the NLO accuracy \cite{Kardos:2013vxa}. It is similar in size (about 
+38\,\% --26\,\% at 14\,TeV and +41\,\% --29\,\% at 8\,TeV, for a jet
\pt\ cut at 40\,GeV) and also uniform in 
shape, supporting our choice for the default scale. 
 
The PDF bands are much narrower. In general, we find that up to 
statistical fluctuations, the smallest cross sections are obtained with 
{\texttt{CT10NLO}}, while {\texttt{MSTW2008NLO}} and {\texttt{NNPDF}} give similar, at most $\sim$ 10\,\% larger values than our default, with {\texttt{MSTW}} predictions slightly larger than {\texttt{NNPDF}} ones. These results show that this complicated final state with rather exclusive experimental selection cuts can be 
modelled reliably in perturbation theory at NLO accuracy matched with 
SMC. 
 
We present an example for lepton distributions in \fig{fig:CMSptl1}.a where 
the spectrum of the transverse momentum of the hardest lepton is depicted. 
This lepton emerges in the decay of one of the t-quarks, i.e.~it is not 
present in the NLO matrix elements. Nevertheless, all what we pointed out 
for the $b$-jet distributions concerning the scale and PDF
uncertainties are valid for this (and other leptonic distributions, not
shown here), too.  

Finally, in \fig{fig:CMSptl1}.b we present the distribution of the 
transverse momentum of the \ttbb\ system, emphasizing the low \pt-region, 
where we see the effect of Sudakov damping. This distribution is 
divergent for $\pt=0$ at fixed order in perturbation theory, while it is 
finite in the matched prediction. We find decreasing scale dependence 
with decreasing \pt, although with naturally worsening statistics. 
 
\section{Conclusions} 
\label{conclusions} 
 
We have studied the \ttbb\ process including NLO QCD corrections 
computed with five massless flavours matched with parton shower, 
using the \powhel\ event generator interfaced with the \pythia\ SMC. 
This computation, challenging due to the presence of two massless b's 
in the final state, producing singular underlying Born configurations, 
opens the road to a unified treatment of \ttbb\ and \ttjj. The latter  
process can also produce \tT+2 $b$-jet final states when both jets are 
tagged as $b$-jets due to $g\to \bB$ splittings. Such contributions are 
not included in our computation. 
 
We presented predictions at different stages of event evolution in 
order to show the effect of the SMC in modifying NLO distributions. 
We found that the effect of the parton shower over the predictions 
at NLO accuracy are usually very small except for rapidity 
distributions and small values of transverse momenta. The effect of 
the hadronization is to soften the transverse momentum or invariant 
mass spectra and can be up to 30\,\%. However, the largest effect is 
due to the decay of the heavy quarks. Kinematic distributions of the 
leptonic decay products of the heavy quarks are not affected by 
the hadronization. 
 
We paid special attention to the identification of the sources of 
uncertainties and in the estimate of the uncertainty bands at the 
hadron level. We found that the scale dependences change only very moderately
with c.m. energy, and are very similar to those of the predictions at the NLO accuracy reported in Ref.~\cite{Kardos:2013vxa}, altough the latter were obtained with looser systems of cuts. These scale dependences 
are fairly uniform for all distributions shown when we employ our default 
scale -- the half of the sum of transverse masses in the final 
state --, which supports our choice.  The PDF 
uncertainties are much smaller, with {\tt CT10NLO} yielding the smallest 
cross section in general, while {\tt MSTW2008NLO} and {\tt NNPDF} giving very 
similar results that are up to about 10\,\% higher.  These results show that 
our events can model reliably the \ttbb\ final states. 

Final states consisting of a \tT\ and a \bB\ quark pair constitute
important backgrounds for Higgs boson production in association with a
\tT-pair. Our events can be used to optimize the selection of the signal
events to find the best signal/background ratio in the various decay
channels. For this purpose sets of LHEs can be downloaded from
our web-page for both the \ttH\ signal 
({\texttt{http://grid.kfki.hu/twiki/bin/view/DbTheory/TthProd}})
and for the background 
({\texttt{http://grid.kfki.hu/twiki/bin/view/DbTheory/TtbbProd}}), 
or new ones can be requested for different choices of scale, PDF,
and parameter values. 

\bigskip\noindent{\bf Acknowledgements}

We are grateful to Malgorzata Worek, Giuseppe Bevilacqua and Tae Jeong Kim for useful discussions on several aspects of their \ttbb\ studies. This research was supported by the Hungarian Scientific Research Fund grant K-101482, the SNF-SCOPES-JRP-2014 grant ``Preparation for and exploitation of the CMS data taking at the next LHC run'', the European Union and the European Social Fund through
LHCPhenoNet network PITN-GA-2010-264564, and the
Supercomputer, the national virtual lab TAMOP-4.2.2.C-11/1/KONV-2012-0010
project.

\bibliographystyle{JHEP} 
%\bibliography{ttbb44} 
\providecommand{\href}[2]{#2}\begingroup\raggedright\endgroup 
\end{document}